# Can biopolymer structures be sampled enumeratively? Atomic-accuracy RNA loop modeling by a stepwise ansatz

Parin Sripakdeevong[1], Wipapat Kladwang[2], Rhiju Das[1,2,3,*]

[1]Biophysics Program, Stanford University, Stanford, CA 94305, USA

[2]Department of Biochemistry, Stanford University, Stanford, CA 94305, USA

[3]Department of Physics, Stanford University, Stanford, CA 94305, USA

[*] To whom correspondence should be addressed. Phone: (650) 723-5976. Fax: (650) 723-6783. E-mail: rhiju@stanford.edu.

## Abstract

Atomic-accuracy structure prediction of macromolecules is a long-sought goal of computational biophysics. Accurate modeling should be achievable by optimizing a physically realistic energy function but is presently precluded by incomplete sampling of a biopolymer's many degrees of freedom. We present herein a working hypothesis, called the "stepwise ansatz", for recursively constructing well-packed atomic-detail models in small steps, enumerating several million conformations for each monomer and covering all build-up paths. By implementing the strategy in Rosetta and making use of high-performance computing, we provide first tests of this hypothesis on a benchmark of fifteen RNA loop modeling problems drawn from riboswitches, ribozymes, and the ribosome, including ten cases that were not solvable by prior knowledge based modeling approaches. For each loop problem, this deterministic stepwise assembly (SWA) method either reaches atomic accuracy or exposes flaws in Rosetta's all-atom energy function, indicating the resolution of the conformational sampling bottleneck. To our knowledge, SWA is the first enumerative, *ab initio* build-up method to systematically outperform existing Monte Carlo and knowledge-based methods for 3D structure prediction. As a rigorous experimental test, we have applied SWA to a small RNA motif of previously unknown structure, the C7.2 tetraloop/tetraloop-receptor, and stringently tested this blind prediction with nucleotide-resolution structure mapping data.

## Introduction

Predicting the three-dimensional structures attained by functional macromolecules is a fundamental challenge in computational biophysics and, more generally, in understanding and engineering living systems. There have been numerous recent successes in the high-resolution modeling of small proteins (1-3), protein/RNA complexes (4), and protein/DNA interfaces (5) by optimizing physically realistic energy functions. Nevertheless, rigorous blind trials demonstrate that the predictive power of computational algorithms remains limited, especially if atomic resolution is sought. For essentially all high-resolution modeling problems tackled to date, the shared critical bottleneck of these methods is inefficient sampling of a biopolymer's vast conformational space (1-7). In addition to hindering accurate modeling, poor sampling precludes rigorous tests of the assumed high-resolution energy functions.

To gain insight into the conformational sampling bottleneck, we have been focusing on some of the smallest well-defined biomolecule folding problems: RNA motifs, as small as four nucleotides in length (8). In addition to offering "toy puzzles" for computational methods (9), these loops, junctions, and tertiary contacts are fundamental building blocks of structured non-coding RNAs; they attain well-defined non-canonical conformations that in turn define the positions of the canonical double helices in three dimensions. A previous study presented a Fragment Assembly with Full-Atom Refinement (FARFAR) method (10), implemented in the Rosetta framework and tested on a benchmark containing 32 RNA motifs with lengths up to 20 nucleotides. While recovering near-atomic-accuracy models in half the cases, FARFAR encountered a conformational sampling bottleneck once motifs exceeded fifteen nucleotides in length. In only one of these longer cases was FARFAR able to generate models with root mean squared deviation (RMSD) less than 1.5 Å to the crystallographic model.

Herein we seek to dissect and to resolve this conformational sampling bottleneck. We focus on an even simpler problem – the structure prediction of irregular RNA loops excised from crystallographic models. As with the analogous protein cases, the RNA loop-modeling problem has important practical significance as a critical component of homology-based structure prediction. In particular, previous methods (RLooM, ModeRNA (11, 12)) are not applicable to loops in homology models that have significantly diverged in sequence or conformation from previously solved structures. As is illustrated below, even the smallest of such loops can be challenging for computational methods, as they are rich in non-canonical nucleotide-nucleotide interactions, bulged nucleotides, and unusual torsion combinations. Solving the RNA loop-modeling problem is a prerequisite for solving the more general RNA structure prediction problem, and we reasoned that a deeper understanding of the conformational sampling bottleneck for loops might be leveraged into general insights for modeling entire functional RNAs or other macromolecules.

Our major finding is that a recursive "ansatz" (hypothesis) enables the systematic sampling of RNA loop conformations at atomic resolution and in polynomial computational time. The method is similar in spirit to *ab initio* "build-up" methods previously explored in protein modeling by Levinthal, Scheraga, Dill, and others (6, 13-15) but abandoned or modified due to their computational cost. Our focus on small RNA loops allows us to revisit and rigorously test these ideas. After illustrating the limitations of knowledge-based approaches on a comprehensive loop-modeling benchmark, we give a complete description of the motivations for the new ansatz, its potential advantages and disadvantages, and its implementation as the stepwise assembly (SWA) method in the Rosetta framework. We then demonstrate substantial improvements in sampling power of the SWA method over prior approaches. In a majority of the cases, the new method enables atomic accuracy structure modeling; and in all the other cases, it reveals previously hidden flaws in the Rosetta all-atom energy function for RNA. As a final rigorous test, we present a blind prediction of an RNA motif of previously unknown structure, the *in vitro* evolved C7.2 receptor for the GAAA tetraloop, along with experimental validation via subsequent chemical accessibility measurements. We end the paper with discussions of historical precedents for this ansatz as well as extensions of this strategy to multi-stranded RNA motifs and protein problems.

## Results

*A benchmark for the RNA loop-modeling problem*

The RNA loop-modeling problem offers small but highly challenging cases for atomic resolution structure prediction. We compiled a benchmark of fifteen loops that begin and end at different Watson-Crick double helices, drawn from riboswitches, ribozymes, and other structured noncoding RNAs with crystallographic data (resolution better than 2.85 Å; see SI Table 1). Loop lengths ranged from 4 to 10 nucleotides (longer loops are rare and occur in <1% of cases; see SI Fig. 1). "Hairpin" loops beginning and ending at the same helix as well as multiple-stranded loops can also be treated but are considered separately (see below).

On one hand, these loops assemble into well-defined conformations, forming a significant number of hydrogen bonds – 2.6 per nucleotide on average, in the same range as values for an A-form RNA helix (2 to 3). For several cases, independent crystallographic structures of the same loop are available and give indistinguishable conformations (SI Table S2). On the other hand, the loops are highly non-canonical. More than half of the hydrogen bonds are in base-phosphate or base-sugar interactions rather than in base pairs (16, 17). Further, the loop torsions are highly irregular. 27% of the nucleotide suites are not part of the 46 most commonly observed RNA rotamers (18); and 9% of the loop nucleotides are bulges. Several loops display sharp turns, as exemplified by the J2/4 loop motif that forms a 140° bend in the 3-way junction of a thiamine pyrophosphate (TPP) sensing riboswitch (Fig. 1A-B). Modeling these intricate loop structures *de novo* is therefore a well-posed but difficult problem.

*Limitations in the knowledge-based methods*

The difficulty of RNA loop modeling is underscored by the poor accuracy of previous methods for RNA structure prediction. For example, a recently developed homology modeling method, RLooM, failed to achieve high-accuracy models in 13 of the 15 benchmark cases, unless directly related loop structures from the same species were permitted (SI Table S2). As a further test, we updated the high-resolution Fragment Assembly of RNA with Full-Atom Refinement (FARFAR) method to permit loop modeling with chain closure and sampling of bulged nucleotides. As before, we mimicked a true modeling scenario by removing any regions with evolutionary kinship to our test motifs.

FARFAR failed to achieve accurate models (under 1.5 Å RMSD to the crystallographic loop, over all loop heavy atoms except crystallographic bulges) as one of the five lowest energy cluster centers in more than half the test cases (11 of 15; Table 1 and SI Table S3). Some of the problem cases are quite small; for example, the J2/4 loop of the TPP riboswitch was not solvable by FARFAR but is only 5 nucleotides in length (Fig. 1A-B). As in prior work, conformational sampling was the dominant bottleneck, as evidenced by three observations. First, for 7 of 11 problem cases, none of the 250,000 models generated gave RMSD accuracy better than 1.5 Å (Table 1). Second, in all cases, this inability to generate near-native structures was traced to the absence of native torsions in the fragment library. The sampling could be rescued by doping native torsions into the fragment library as a "cheat" to aid conformational search (see SI Supporting Methods & SI Table 4). Third, in 9 of 11 cases, the generated models did not achieve near-native energies; the lowest energy of 250,000 models remained higher than the energy of optimized native loops (Table 1). The inability of FARFAR to solve these small loop modeling problems suggests that one or more basic assumptions of the fragment assembly approach limits its conformational sampling power.

*A stepwise ansatz*

We reasoned that the conformations of small loops might be effectively sampled through direct enumeration at high resolution, rather than by restricting the search space to previously known fragments. We discovered that a recursive step-by-step enumeration (Figs. 1C-F) permits efficient *de novo* sampling of these problems, which we illustrate on one of the FARFAR failures above, the J2/4 loop of the TPP riboswitch.

First, we note that exhaustive enumeration of this five-nucleotide loop at atomic resolution is not feasible with current computational power. Even building one nucleotide of a loop involves sampling several degrees of freedom, including 6 backbone torsions, 4 (coupled) sugar-pucker torsions, the glycosidic torsion, and the 2′-OH torsion. While low-resolution (>3 Å) clustering of exhaustively sampled single-nucleotide conformations results in under 100 "rotamers" [SI Table 5; see also (18)], high-resolution clustering with a sub-Angstrom threshold leads to millions of unique conformations of the nucleotide (SI Table 5). While modeling and computing the Rosetta energy for this number of conformations is achievable in less than 1 hour on a single modern CPU, the available conformations multiply exponentially with the RNA length. Thus, combinatorial enumeration of all available conformations of a five-nucleotide loop would require ~$10^{23}$ CPU-years, well beyond the computational power achievable in the foreseeable future.

Nevertheless, the feasibility of enumerating the conformations of just one nucleotide suggests an alternative approach to realistic RNA modeling. Enumerative single-nucleotide building permits fine-grained exploration of torsional conformations that form well-packed structures with multiple hydrogen bonds, as is observed in native loops. As an illustration, Fig. 1C shows the lowest energy conformation in the J2/4 loop achieved by exhaustive sampling, followed by local energy minimization, of the first nucleotide built from the 3′ end. The resulting nucleotide is positioned with atomic accuracy, giving an RMSD of 0.69 Å from the experimental conformation. We discovered that the entire loop could then be recovered through stepwise building of each additional nucleotide (Fig. 1C-F), carrying forward an ensemble of the lowest energy well-packed, well-hydrogen-bonded conformations from each previous sub-region. In addition to standard single-nucleotide building steps, regenerating this loop also required a "bulge-skip" building step (to permit the modeling of unpaired/unstacked nucleotides) and a chain-closure building step to complete the RNA loop (e.g., Fig. 1E-F; see SI Supporting Methods for complete descriptions of the three types of building steps).

The remaining challenges in this step-by-step approach are the choice of the building pathways and the choice of models to carry forward. In particular, we do not know *a priori* the appropriate order of building steps that will achieve the experimental conformation, and we cannot guarantee that the lowest energy model for a sub-region will carry forward into the lowest energy model for the entire loop. Further, the number of such build-up paths grows exponentially with the number of nucleotides. We solved these path-enumeration issues using a recursive strategy, familiar from dynamic programming approaches applied to sequence alignment (19) and to RNA secondary structure prediction (20, 21). We determined a low energy ensemble of models for each sub-region of the loop as modeled from the 5′ end or from the 3′ end and then join all combinations of these sub-regions by chain closure. A directed acyclic graph (22) delineates this overall recursive calculation, as shown in Fig. 1G. Briefly, a model for each sub-region was obtained in one of two ways – from either a standard single-nucleotide building step from a sub-region one nucleotide shorter, or from a bulge-skip two-nucleotide building step from a sub-region two nucleotides shorter. All models for a sub-region were clustered, and the thousand lowest energy cluster centers (which typically included all models within 6 $k_{\text{B}}T$ of the lowest energy state, mimicking models accessed by thermal fluctuations) were carried forward. In the case of the J2/4 loop example, searching through all possible paths led to a diverse set of well-packed conformations, including low-energy near-native and non-native models that were completely missed by FARFAR (Fig. 1H).

This stepwise assembly method deterministically enumerates a low-energy subspace of the RNA loop conformational space and guarantees that each nucleotide forms well-optimized interactions with nucleotides outside the loop or previously built in the loop. Nevertheless, the method cannot guarantee, *a priori*, the sampling of the globally lowest energy conformation. The method instead assumes, as a working hypothesis, that the experimentally observed conformation is achievable through the stepwise, locally optimal building of individual nucleotides. This stepwise ansatz can only be confirmed through empirical tests on naturally occurring macromolecule sequences. We have therefore carried out extensive trials of this working hypothesis using RNA loop modeling as a biophysically important but unsolved test problem, described next.

*Comprehensive test of the stepwise ansatz*

To evaluate the validity of the stepwise ansatz, we applied it to the entire 15-loop benchmark (Table S1). In terms of modeling accuracy, SWA substantially outperformed FARFAR, recovering near-native models (< 1.5 Å RMSD) for 10 of 15 test cases, compared to 4 cases recovered by FARFAR (see Table 1). These included atomic-accuracy models from diverse sources, including a 5-nt loop from the J5/5a hinge in the P4-P6 domain of the group I *Tetrahymena* ribozyme (RMSD of 1.04 Å; Fig. 2A); a 7-nt loop connecting helices P2 and P3 of the group II intron (RMSD of 1.25 Å; Fig. 2B); and one of the two 10-nt loops in the benchmark, nucleotides 2003-2012 of the large ribosomal subunit from *H.*

*marismortui* (RMSD of 0.74 Å; Fig. 2C). In each of these 10 success cases, the high accuracy of the SWA model is reflected not only in low RMSD to the experimental loops (<1.5 Å) but also near-complete recovery of the base pair geometries as classified in the Leontis-Westhof scheme (16) (SI Table S6).

For the remaining five "problem cases", conformational sampling was no longer the major bottleneck. In all five cases, SWA models achieved lower energies than optimized native models (see Table 1 & Fig. 3A). Further, in four of the five cases, SWA sampled *de novo* models within 1.5 Å of the experimental conformation, although these models were not selected as one of the five lowest energy cluster centers. In the last case (a second 10-nt ribosomal loop), the optimized native model gave significantly worse energy (by 7.5 $k_B T$) than the SWA models, explaining the absence of near-native models in the low-energy SWA ensemble. These results demonstrated that the stepwise ansatz is valid in all tested cases, and inaccuracies in the lowest energy models were due to approximations or omissions in the Rosetta all-atom energy function. The results were in strong contrast to the FARFAR results above.

*Blind Prediction and Experimental Validation*

The most stringent tests for structure prediction algorithms are blind trials – modeling of molecules with no known experimental structure. There have been few attempts at blind high-resolution RNA structure modeling, and so far these have not achieved atomic accuracy [see, e.g., (23-27)]. Encouraged by the strong performance of SWA on the benchmark, we predicted the structure of a novel tetraloop receptor motif (the C7.2 mutant; Fig. 4A) with no known experimental structure (28, 29), previously isolated by *in vitro* selection on the *td* self-splicing group I intron.

This sequence served as an appropriate first blind test because it effectively reduces to a small but challenging loop-modeling problem. Much of the sequence aligns with a widely studied tetraloop receptor motif whose structure has been determined by crystallography in several different RNAs, including the P4-P6 domain of the *Tetrahymena* ribozyme (30, 31). The main difference is a three-nucleotide segment (G4-U5-A6) replacing a two-nucleotide A4-A5 "platform". We modeled this segment by SWA, FARFAR, RLooM, and ModeRNA. SWA gave the well-packed C7.2 receptor model shown in Fig. 4B as the lowest energy structure. More extensive SWA calculations modeling eight nucleotides (nucleotides 3–7 and 10–12 in Fig. 4A) gave similar structures. FARFAR predicts a model that is different from the SWA conformation and is worse in energy (by 3 $k_B T$; SI Fig. S2B); the presence of several different FARFAR models with similar energies also precluded confident modeling. The RLooM and ModeRNA models contained numerous steric clashes as identified by MolProbity (32) as well as empty cavities in the core of the structures (SI Fig. S2E-F). Thus, despite its small size, the C7.2 three-nucleotide segment was a challenging case for high-resolution modeling.

The predicted SWA model for the C7.2 receptor displayed unusual features, as compared to the A-A platform of the classic tetraloop receptor. First, the central U5 nucleotide bulged out of the structure. Second, the first and third nucleotides of the loop formed a same-stranded sugar-edge/Watson-Crick G4-A6 base pair (Fig. 4B) separated by the bulged nucleotide. The FR3D motif search software (33) found only two other instances of this conformation in the entire database of RNA structures, within a malachite green aptamer (PDB: 1F1T) and in a complex of the UUGUAU RNA to the human cleavage factor protein Im (PDB: 3MDG). Nevertheless, the neighboring 5′ and 3′ nucleotides in these two precedent structures are positioned differently than in the tetraloop receptor (FR3D geometric discrepancies of 0.72 Å and 0.89 Å; both worse than the default cutoff 0.5 Å), explaining the inability of RLooM and ModeRNA to produce reasonable models.

The SWA model for the C7.2 receptor (Fig. 4B) makes predictions for the motif's accessibilities that can be tested by single-nucleotide-resolution chemical modification experiments. Nucleotides outside the G4-U5-A6 loop should give modification rates similar to measurements in the classic tetraloop receptor motif. The reagent dimethyl sulfate (DMS) alkylates A or C nucleotides whose Watson-Crick edges are not paired (see, e.g., (34)); the SWA model predicts that A6 should be protected from DMS modification. The carbodiimide reagent CMCT modifies the Watson-Crick face of G or U nucleotides that are unpaired and unstacked (35, 36); the SWA model predicts that the bulged U5 should be highly modified by CMCT but the well-stacked G4 base should be protected. Other models (SI Fig. S2) give different predictions; e.g., A4 would be exposed (FARFAR & RLooM) or U5 would be protected (ModeRNA).

We tested these blind predictions by grafting the C7.2 receptor into the J6a/6b and J6b/6a segments of the P4-P6 RNA (SI Fig. S3A) and carrying out quantitative chemical mapping, read out by single-nucleotide-resolution capillary electrophoresis. As with the wild type P4-P6 RNA, the C7.2-grafted variant showed strong protections of the L5b tetraloop, J6a/6b tetraloop receptor, and the P5a A-rich bulge upon addition of $Mg^{2+}$, verifying the attainment of the RNA's global tertiary fold (electrophorograms shown in SI Fig. S4). Further, as expected, we observed indistinguishable modification rates between the wild type RNA and the C7.2 variant outside the tetraloop receptor (SI Figs. S3B-D). Focusing on the substituted C7.2 receptor, nucleotides G4 and A6 were both protected from chemical modification, as predicted in the SWA model (nts 225 & 227 in conventional P4-P6 numbering; Fig. 4C & SI Figs. S3). Most importantly, U5 (nt 226 in conventional numbering) was highly modified by CMCT, at a rate 19±1 times the mean modification rate for Watson-Crick base-paired uridines in the adjoining P6b helix (U228, U241, U243, and U244 in conventional numbering). As positive controls, the entire P4-P6 RNA molecule gave six other uridines that achieved CMCT modification rates of at least 10-fold above background (U130, U179, U185, U199, U236, and U238 in conventional numbering). All of these nucleotides are bulged in crystallographic models of the P4-P6 RNA, providing additional confirmation that U5 in the mutant C7.2 receptor is indeed bulged, as predicted. The chemical accessibility data thus validate the *de novo* SWA model at nucleotide resolution, and strongly disfavor models from knowledge-based methods (SI Fig. S2).

After obtaining these experimental results, we discovered further evidence for the SWA model from sequence variations in the original *in vitro* experiment that isolated receptor subclass C7.2 (28). The predicted G4-A6 base pair is only isosteric to one other base pair (A-A), which has only a single polar hydrogen bond, (33) so we expected to see few if any substitutions. The predicted U5 bulge is unpaired and thus should be replaceable by other nucleotides. Indeed, in all sequenced clones, G4 and A6 remained invariant, while mutations of U5 to A and C were observed, giving further validation of the SWA model. Further atomic-resolution tests might be achieved if crystals can be obtained for the P4-P6 RNA C7.2 variant; crystallization trials are under way but incomplete (personal communication, E. Leung, J. Piccirilli). Community-wide blind trials of high resolution RNA structure modeling (personal communication, E. Westhof, N. Leontis) should provide further rigorous tests for the SWA approach.

## Discussion

*A stepwise ansatz resolves a conformational sampling bottleneck in structure prediction*

An inability to guarantee exhaustive conformational sampling prevents high-resolution biomolecule structure prediction methods from attaining wide acceptance and usage. In Rosetta as well as other frameworks such as MC-Fold/MC-Sym (37), methods for protein and nucleic acid modeling typically make use of knowledge-based search in a reduced representation followed by high-resolution refinement. As has been discussed previously (9, 38), potential issues hampering these methods' *de novo* sampling efficiency include their dependence on the database of existing experimental structures; the stochasticity of Monte Carlo fragment assembly; and the use of coarse-grained phases to smooth and reduce the dimensionality of the search space.

To address these issues, we developed a working hypothesis, called the "stepwise ansatz", and its implementation, the stepwise assembly method, that enumeratively searches a physically realistic subspace of a molecule's all-atom conformations in polynomial computational time [$O(N)$ where $N$ is the number of nucleotides]. The concept of *ab initio* step-by-step build-up has been discussed previously, e.g., in enumerative coarse-grained or stochastic all-atom search methods from Dill & colleagues (15, 39-41), pioneering peptide-modeling work from the 1980s by the Scheraga lab (13, 14), and earlier computational explorations by Levinthal in 1968 (6). However, these prior build-up strategies have not been adopted into the mainstream of structure modeling or shown to outcompete Monte Carlo or knowledge-based methods (41). The prior lack of development appears to stem from the difficulty of searching all possible build-up paths and from the high expense of the deterministic enumerative calculations relative to stochastic, knowledge-based methods. For example, modeling a single 5-nucleotide RNA loop herein required 12,000 CPU-hours; fortunately, this calculation is now feasible due to the growing availability, decreasing cost, and massive parallelization of high-performance computer clusters. This computational power and our technical implementation allows the first rigorous test of a biopolymer structure prediction method that is independent of the databases of existing structures, fundamentally deterministic (rather than stochastic), and free from coarse-grained or reduced representations while optimizing an all-atom energy function.

On a challenging benchmark of irregular RNA loop motifs, we have shown that SWA resolves the conformational sampling bottleneck that has hindered prior knowledge-based methods. In all cases, SWA either sampled the experimental loop conformation *de novo* or recovered conformations with energies that surpassed the energy of the optimized experimental loop conformation. These results confirm the underlying ansatz that low-energy full-length conformations can be reached through step-by-step, enumerative building of individual nucleotides, with all building paths sampled through recursion. Further, in the majority of the cases (10 of 15), the Rosetta all-atom energy function was accurate enough to permit a near-native conformation to be selected as one of the five lowest energy cluster centers. The strongest test of the SWA method is the blind prediction on the C7.2 tetraloop-receptor motif of presently unknown structure. The predicted model includes non-canonical features (including a same-stranded G-A base pair and a bulged nucleotide) and agrees with subsequently measured chemical accessibility data; more stringent tests of the model are being attempted through x-ray crystallography.

*Inaccuracies revealed in the high-resolution Rosetta energy function*

Prior studies have reported anecdotal cases of failures of the high-resolution Rosetta energy function for macromolecule modeling (9, 42), but the work herein is the first example of a complete high-resolution *de novo* modeling benchmark in which effective conformational sampling has been demonstrated for every test case. Looking at the top five clusters, the overall rate of successful discrimination by this energy function is reasonably high (10 out of 15). Nevertheless, in only 5 of these cases is the very lowest energy model (rather than just one of the five lowest energy cluster centers) in atomic-resolution agreement with the experimental conformation. Further, in 4 of the 15 cases, optimized experimental conformations are at least 4 $k_BT$ worse in energy than the lowest-energy models. These results unambiguously indicate that approximations in the Rosetta all-atom energy function remain too inaccurate to permit atomic-resolution RNA modeling on a consistent basis. The energy function does not explicitly model metal ions (as are observed in crystals of the hepatitis C virus IRES subdomain IIa (43); Fig. 3A), and water is modeled through a crude solvation term (44). Collective dispersion effects, relevant for interactions of large groups like nucleobases; long-range electrostatic effects; and hydrogen bond cooperativity are presently neglected. Furthermore, conformational entropies for each model (along with definitions of associated conformational ensembles) are not calculated. Nevertheless, given the enumerative nature of SWA, modifications to the code should allow for a practical calculation of the conformational entropy. Further, the generality of the search method, and the small size of the RNA loop modeling puzzles, should permit strong tests of more recently developed all-atom energy functions, including those that model polarizable moieties (45). Finally, a new generation of structure determination methods uses limited experimental information (e.g., backbone chemical shifts) to break degeneracies in existing energy functions [see, e.g., (46-48)]; the SWA approach should allow optimization of these hybrid *de novo*/experimental problems with unprecedented efficiency.

*A potentially general strategy for molecular modeling*

In this work, we have focused mainly on the application of SWA toward single-stranded RNA loop-segments, both to demonstrate the method's conformational sampling power and to solve a basic practical problem that arises in RNA structure prediction. Nevertheless, the basic building blocks of RNA tertiary structure, non-canonical motifs, often involve multiple RNA strands interacting with one another or strands returning to the same helix. Extensions of the basic SWA method to predict these motifs' structures by recursively modeling sub-regions appears accurate, and the calculation still grows in polynomial time with the number of nucleotides (unpub. data, PS, RD; see SI Fig. S5). With expected improvements in computational power and the method's efficiency, *de novo* high-resolution modeling of RNA motifs with lengths up to 15 nucleotides, a size range that includes many RNA aptamers and catalytic sites, should be feasible. Further, the basic concepts underlying stepwise assembly are not specific to RNA. We look forward to testing this conformational search strategy in other frontier problems in high-resolution macromolecule modeling, including efficient prediction of protein loops and small proteins, rigorous tests of protein and protein/RNA energy functions, and enumerative sequence design of functional protein and RNA loops.

**Acknowledgments**

We are grateful to the Das lab for comments on the manuscript and members of the Rosetta community for discussions and sharing of code. Calculations were carried out on the BioX$^2$ cluster (NSF award CNS-0619926). We acknowledge financial support from a Burroughs-Wellcome Career Award at the Scientific Interface (to RD) and a C. V. Starr Asia/Pacific Stanford Graduate Fellowship (to PS).

**Methods**

*Code implementation in Rosetta*

The stepwise assembly method proceeds through the recursive building of single nucleotides. Complete descriptions of the general scheme of the recursion, and the implementation of the single-nucleotide building step, the bulge-skip building step, the chain-closure step, and the models clustering step used in the SWA method are provided in SI Supporting Methods. The FARFAR method was also updated to permit loop modeling (see SI Supporting Methods). A strategy to parse out loop modeling problems from large structures (such as the large ribosomal subunit), a procedure to generate optimized native models, improvements of the all-atom Rosetta energy function, and explicit command-line examples are also provided in SI Supporting Methods.

*Code implementation and availability*

The SWA and FARFAR methods were implemented in C++ in the Rosetta codebase. Python scripts automatically set up and controlled the workflow of the entire SWA calculation, permitting efficient queuing of the computation on high-performance Linux clusters using either the Load Sharing Facility or Condor queuing systems (see SI Supporting Methods). The software is being made available in the next Rosetta release (3.4), which will be freely available to academic users at http://www.rosettacommons.org. Prior to release, this code will be gladly provided to academic users upon request.

*Prior methods: RLooM, ModeRNA*

RLooM (database version 12-19-08) and ModeRNA (version 1.6.0) were applied to loop modeling problems following instructions given in the released software. Explicit command-line examples used for both packages are given in SI Supporting Methods.

*RNA structure mapping through chemical modification*

DMS and CMCT modification data of the wild type P4-P6 RNA as well as the variant with the C7.2 tetraloop receptor (full sequences in SI Supporting Methods) were acquired at single-nucleotide resolution, as described previously (49). Briefly, preparation of DNA templates, *in vitro* transcription of RNAs, DMS and CMCT chemical mapping, and capillary electrophoresis were carried out in 96-well format, accelerated through the use of magnetic bead purification steps, as has been described previously. Data were analyzed with the HiTRACE version of the SAFA software package (50); background subtraction and correction for attenuation of reverse transcription products were carried out as in (49); and figures prepared in MATLAB.

**Table 1.** Accuracy and conformational sampling efficiency of *de novo* RNA loop modeling

| Motif name | Motif properties | | Best RMSD[a] (Å) of 5 lowest energy clusters[b] | | Lowest RMSD[a] (Å) achieved | | Energy gap to optimize native[c] | |
|---|---|---|---|---|---|---|---|---|
| | Length | PDB | FARFAR | SWA | FARFAR | SWA | FARFAR | SWA |
| 5´ J1/2, Leadzyme | 4 | 1NUJ | 1.96 | **0.83** | 1.66 | **0.51** | 2.7 | **−0.8** |
| 5´ P1, M-Box Riboswitch | 4 | 2QBZ | **0.72** | **0.96** | **0.53** | 0.61 | 2.3 | **−0.5** |
| 3´ J5/5a, Group 1 Intron | 4 | 2R8S | **0.40** | **0.47** | **0.30** | **0.40** | 0.0 | 0.0 |
| 5´ J5/5a, Group 1 Intron | 5 | 2R8S | 4.08 | **1.04** | **1.05** | **0.66** | 0.3 | **−0.9** |
| Hepatitis C Virus IRES IIa | 5 | 2PN4 | 2.11 | 5.31 | **1.04** | **0.71** | −2.6 | **−5.9** |
| J2/4, TPP Riboswitch | 5 | 3D2V | 6.66 | **0.85** | 1.74 | **0.73** | 10.8 | **−1.0** |
| 23S rRNA (44-49) | 6 | 1S72 | **0.69** | **0.73** | **0.47** | **0.71** | 2.6 | 0.0 |
| 23S rRNA (531-536) | 6 | 1S72 | 3.18 | 2.45 | 2.44 | **0.76** | 6.9 | **−0.6** |
| J3/1, Glycine Riboswitch | 7 | 3OWI | **1.13** | **1.35** | **0.71** | **0.64** | 2.5 | 1.3 |
| J2/3, Group II Intron | 7 | 3G78 | 1.72 | **1.25** | 1.53 | **0.96** | 8.5 | **−0.2** |
| L1, SAM-II Riboswitch | 7 | 2QWY | 2.43 | **1.26** | **1.43** | **0.86** | 3.8 | **−1.3** |
| L2, Viral RNA Pseudoknot | 7 | 1L2X | 5.44 | 3.36 | **1.35** | **0.91** | 3.7 | **−4.1** |
| 23S rRNA (2534-2540) | 7 | 1S72 | 6.39 | 5.71 | 3.24 | **1.39** | 7.3 | **−7.3** |
| 23S rRNA (1976-1985) | 10 | 1S72 | 11.54 | 7.92 | 5.08 | 4.42 | 13.0 | **−7.5** |
| 23S rRNA (2003-2012) | 10 | 1S72 | 11.36 | **0.74** | 5.43 | **0.64** | 41.2 | 3.2 |
| **RMSD < 1.50 Å** | – | – | 4/15 | 10/15 | 8/15 | 14/15 | – | – |
| **Energy Gap < 0.0** | – | – | – | – | – | – | 2/15 | 13/15 |

[a] All-heavy-atom RMSD to the crystallographic loop. Nucleotides found to be bulged (both unpaired and unstacked) in the native crystallographic model were excluded from the RMSD calculation. Bold text indicates RMSD is within 1.5 Å of the crystal structure.

[b] Generated models were clustered, such that models with pairwise all-heavy-atom RMSD less than 1.5 Å over entire loop and less than 2.5 Å over each individual loop nucleotide grouped. The lowest energy members of the five lowest energy clusters were designated as the predicted models.

[c] Bold text indicates that the lowest energy sampled by the *denovo* run is lower than the energy of the optimized native (i.e. the energy gap is negative). 1 Rosetta Unit (RU) is approximate equal to $1k_BT$ (10, 51).

**Figure Legends**

**Figure 1. The** **stepwise** **assembly (SWA) structure modeling method**. Illustration on the J2/4 loop segment from the 3-way junction of a thiamine pyrophosphate (TPP) sensing riboswitch (PDB: 3DV2). (A) Crystallographic conformation of the five-nucleotide loop (shown in color) with surrounding nucleotides from the crystallographic model (shown in white). (B) Schematic of the 3-way junction in the annotation of Leontis and Westhof (16); only nucleotides shown in the 3D structure are numbered. (C-F) A build-up path that leads to the experimental conformation; the 5 nucleotides in the loop are built in a stepwise manner, one at time, starting from the 3´ end. (G) A directed acyclic graph delineates the enumerative steps in the SWA method, recursively covering all possible build-up paths. (H) Rosetta all-atom energy vs. all-heavy-atom root mean squared deviation (RMSD) to the crystallographic structure for *de novo* models generated by SWA (blue points) and by the prior method (Fragment Assembly of RNA with Full-Atom Refinement, FARFAR, red points).

**Figure 2. Comparison of crystallographic model and SWA *de novo* models for three diverse loop motifs.** (A) Five-nucleotide loop from the J5/5a hinge in the P4-P6 domain of the group I *Tetrahymena* ribozyme (PDB: 2R8S). (B) Seven-nucleotide loop connecting helices P2 and P3 of the group II intron (PDB: 3G78). (C) Ten-nucleotide loops from the large ribosomal subunit from *H. marismortui* (PDB: 1S72, nts 2003-2012). The modeled loop segment is shown in color while the surrounding crystallographic model is shown in white. Some surrounding nucleotides in the structure are also omitted to permit unobstructed view of the loop-segment region. The RMSDs to the crystallographic conformations (energy cluster rank) of the displayed models were (A) 1.04 Å (fourth), (B) 1.25 Å (fourth), and (C) 0.74 Å (second).

**Figure 3. Limitations in modeling accuracy are no longer due to conformational sampling.** Modeling of a five-nucleotide loop from the hepatitis C virus IRES subdomain IIa. (A) Crystallographic model (PDB: 2PN4). Bound divalent $Sr^{2+}$ cations (colored yellow) are proposed to stabilize the loop through both direct hydrogen bonds as well as through hydrogen bonds mediated by tightly bound water molecules (O atoms colored red) (43). (B) SWA successfully samples the native conformation of the loop to atomic accuracy (0.71 Å all-heavy-atom RMSD), but the model (C) is not ranked within one of the five

lowest energy clusters. (D) The Rosetta all-atom energy function incorrectly assigns numerous non-native models (>5.0 Å RMSD) with significantly lower energies (~6 $k_BT$) than the optimized native structure (red point in B).

**Figure 4. Blind prediction of the C7.2 tetraloop receptor and validation through single-nucleotide-resolution chemical mapping.** (A) Secondary structure model of the C7.2 tetraloop receptor; the 3-nt GUA loop-segment at the core of the receptor (shown in color) is different from receptors with previously solved structures. (B) Three-dimensional model of the C7.2 receptor by stepwise assembly. Models from other methods are given in SI Fig. S2. (C) Chemical reactivities of A and C (based on dimethyl sulfate alkylation) and G and U (based on CMCT carbodiimide modification) shown as white-to-red coloring on a mutant of the P4-P6 domain of the *Tetrahymena* ribozyme containing the C7.2 receptor; measurements were acquired in 10 mM MgCl₂, 50 mM HEPES, pH 8.0, at 24 °C. (D) Bar graph of reactivities for nucleotides near the C7.2 receptor. See SI Figs. S3 & S4 for full data sets, including both wild type and C7.2 variant data and error analysis.

# Figure 1

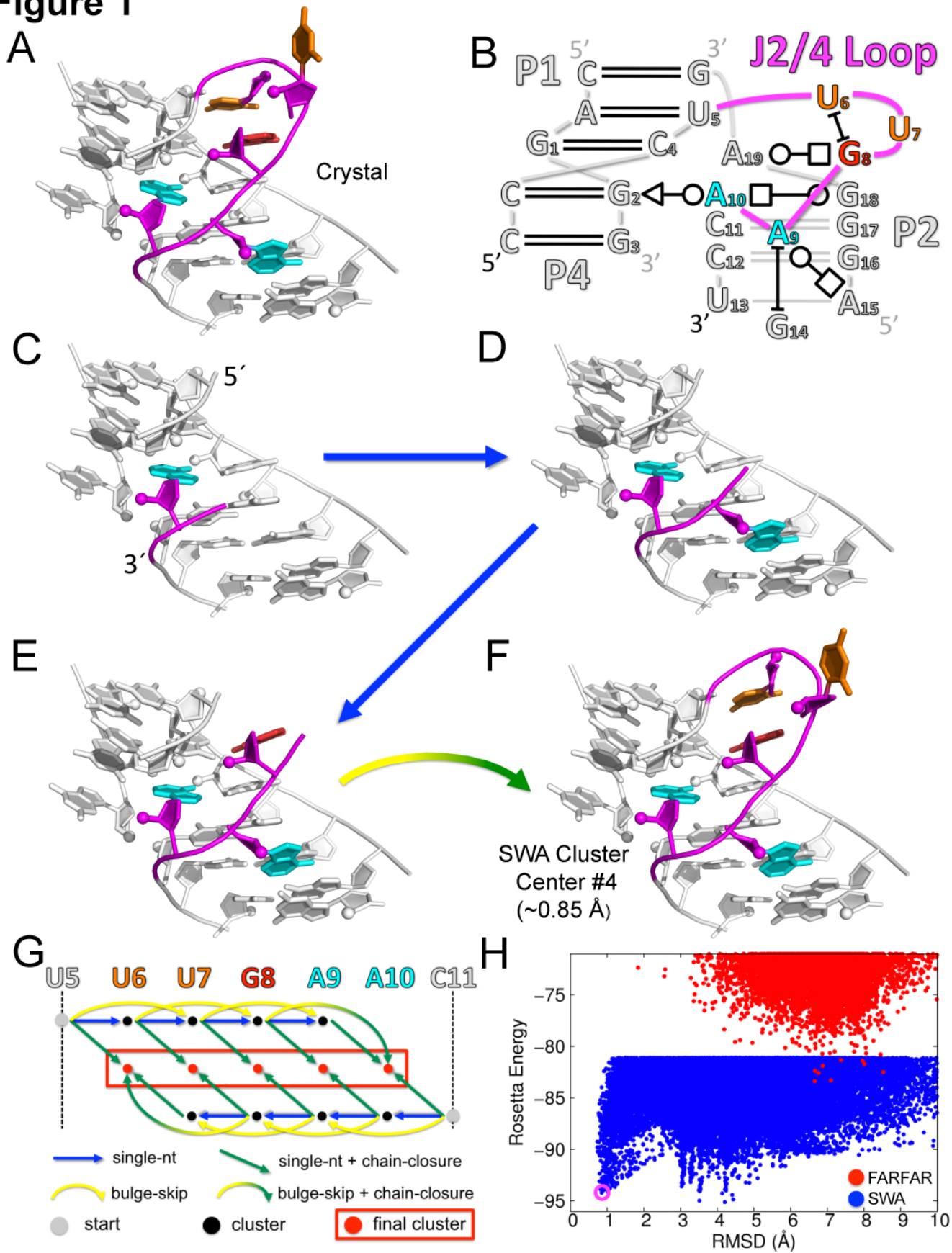

A. Crystal

B. P1, J2/4 Loop, P4, P2

C. 5′ 3′

D.

E.

F. SWA Cluster Center #4 (~0.85 Å)

G. U5 U6 U7 G8 A9 A10 C11

- → single-nt
- ⇒ single-nt + chain-closure
- ⤳ bulge-skip
- ⤳ bulge-skip + chain-closure
- ○ start
- ● cluster
- ● final cluster

H. FARFAR (red) SWA (blue)
Rosetta Energy vs. RMSD (Å)

**Figure 2**

Crystal          SWA Model

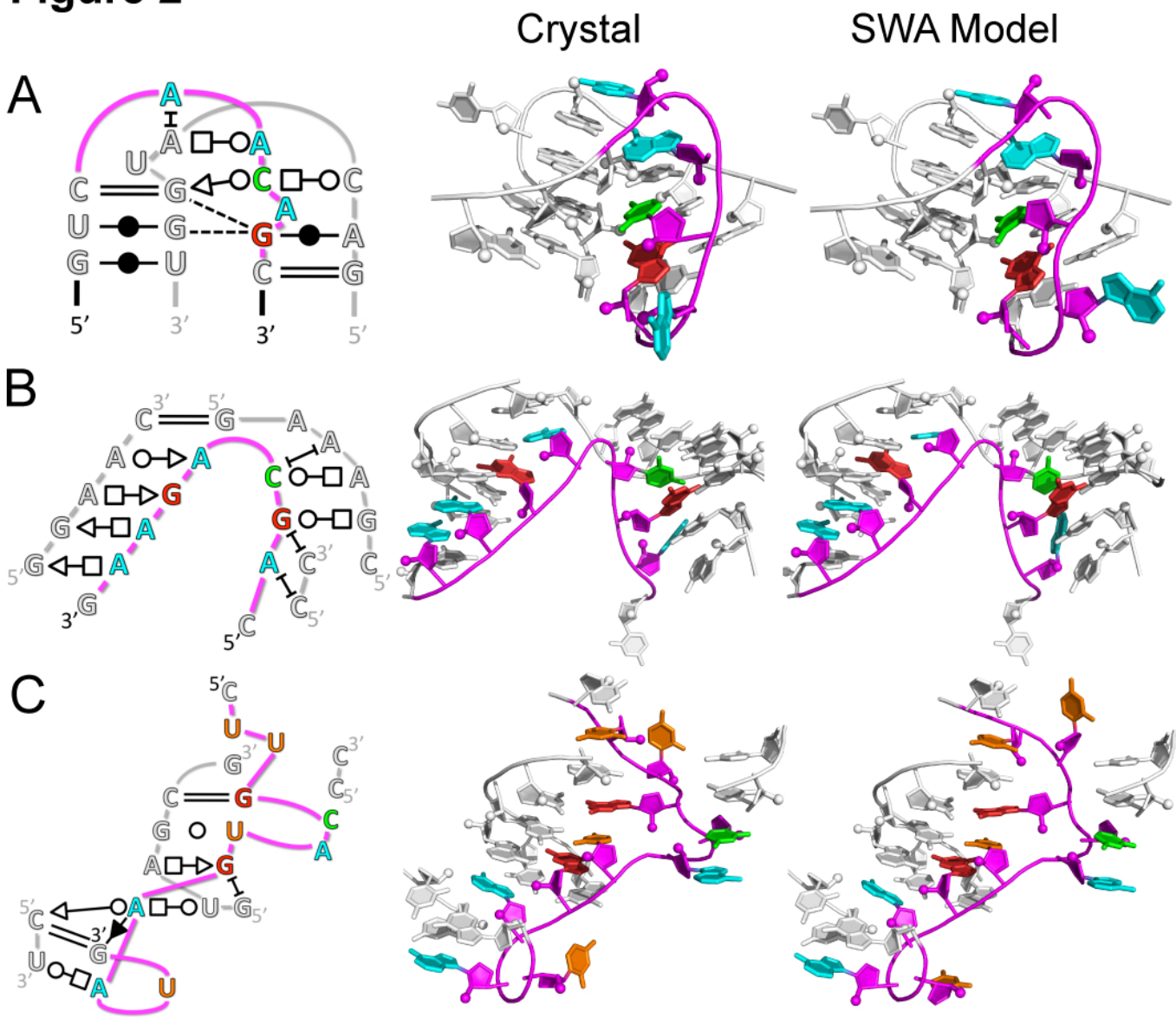

A

B

C

**Figure 3**

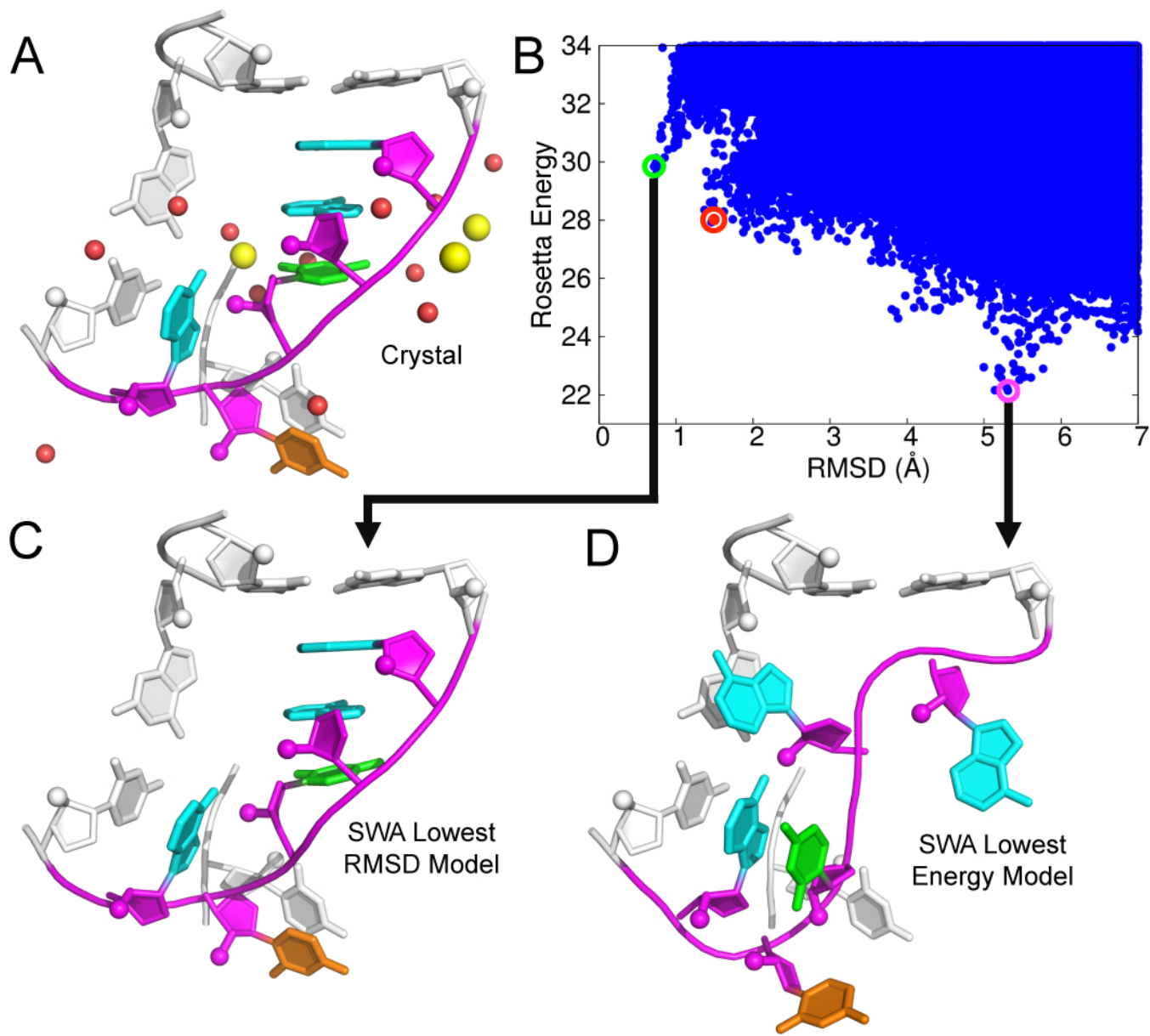

A — Crystal

B — Rosetta Energy vs. RMSD (Å)

C — SWA Lowest RMSD Model

D — SWA Lowest Energy Model

## Figure 4

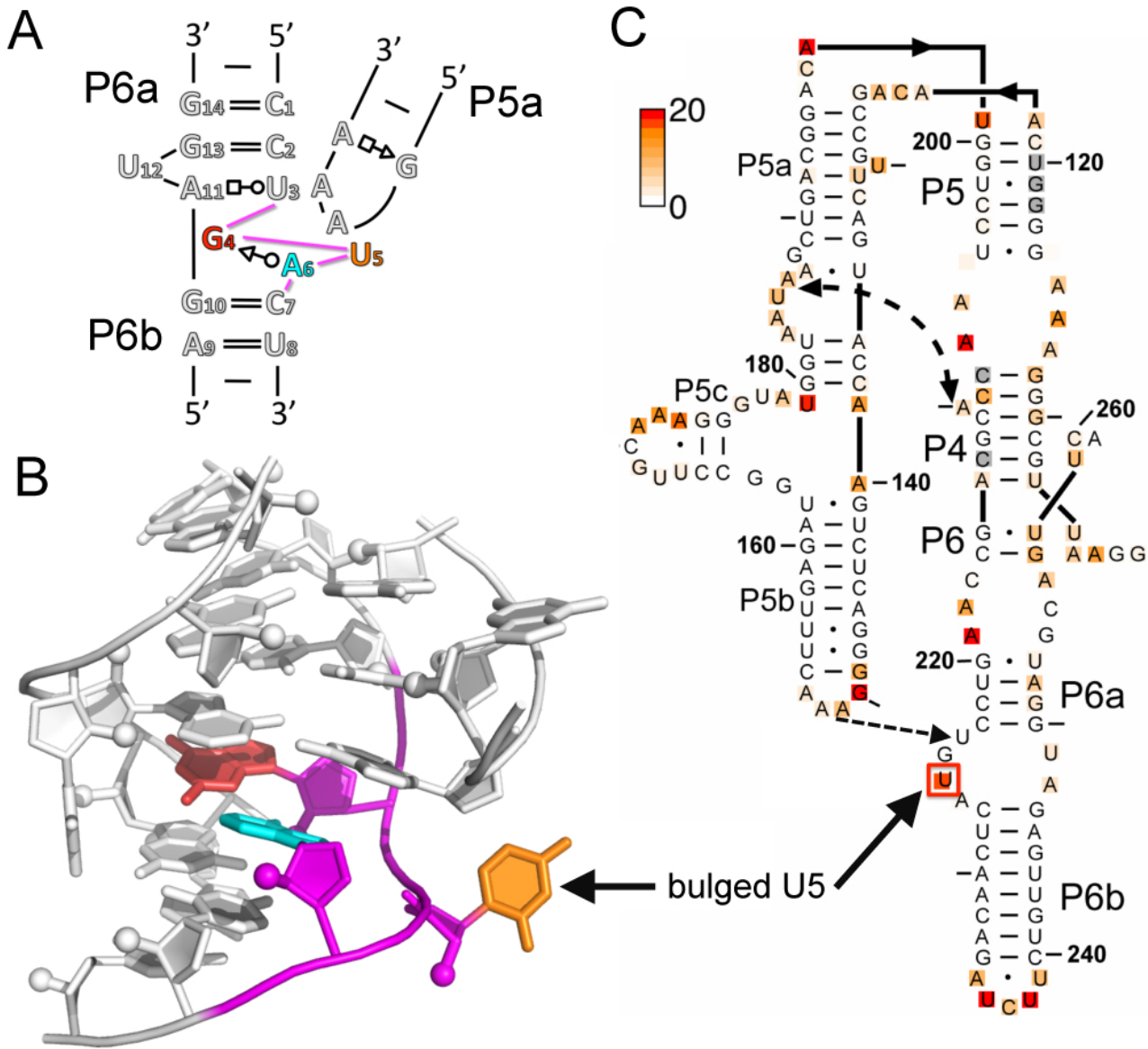

**Supporting Information** for "Can biopolymer structures be sampled enumeratively? Atomic-accuracy RNA loop modeling by a stepwise ansatz"

Parin Sripakdeevong[1], Wipapat Kladwang[2], Rhiju Das[1,2,3,*]

[1]Biophysics Program, Stanford University, Stanford, CA 94305, USA

[2]Department of Biochemistry, Stanford University, Stanford, CA 94305, USA

[3]Department of Physics, Stanford University, Stanford, CA 94305, USA

[*] To whom correspondence should be addressed. Phone: (650) 723-5976. Fax: (650) 723-6783. E-mail: rhiju@stanford.edu.

**List of Supporting Text sections:**

**1. Supporting Methods.**

**2. Supporting Results.**

**List of Supporting Tables:**

**Table S1.** Loop Motifs Benchmark Information.

**Table S2.** Application of RLooM homology modeling to the loop motifs benchmark.

**Table S3.** Supplemental loop motifs benchmark data (FARFAR).

**Table S4.** Comparison of FARFAR with and without doping native fragments into the fragment library.

**Table S5.** Number of conformation cluster centers in a nucleotide versus the RMSD cluster radius.

**Table S6.** Supplemental loop motifs benchmark data (SWA).

**List of Supporting Figures:**

**Figure S1.** RNA loop length distribution in 23S rRNA.

**Figure S2.** Modeling the C7.2 Tetraloop Receptor with RLooM and ModeRNA.

**Figure S3.** Chemical reactivity of the bound C7.2 tetraloop receptor, grafted into the P4-P6 RNA.

**Figure S4.** Capillary electropherograms of chemical mapping experiments on the wild type and C7.2-substituted P4-P6 RNAs.

**Figure S5.** Extension of the SWA method to treat hairpins and multiple-stranded loops.

## Supporting Methods

***Stepwise Assembly in Rosetta***

The Stepwise Assembly method proceeds through the recursive building of single nucleotides. This section describes (a) the general recursive scheme, (b) the single-nucleotide building step, (c) the bulge-skip building step, (d) the chain-closure step, and (e) the models clustering step used in the method.

*(a) General recursive scheme*

The Stepwise Assembly method determines low-energy models for each sub-region of the loop, culminating in models of the entire loop, through recursive application of four kinds of steps (see Fig. 1G). Given the loop nucleotides $i$ through $j$ [denoted here $(i, j)$], the sub-regions consisted of loop pieces built from the 5′ end $(i, k)$, with $i \leq k \leq j$, and loop pieces built from the 3′ end $(l, j)$ with $i \leq l \leq j$. The recursion was defined as follows:

i.      For a 5′-loop piece $(i, k)$, models were generated by applying the *single-nucleotide building step* on models available for the smaller loop piece $(i, k–1)$. Similarly, for a 3′-loop piece $(l, j)$, models were built from the models of the smaller loop piece $(l+1, j)$. The recursion proceeds until the loop is completed [i.e, the full length loop $(i, j)$].

ii.     An additional set of models for each loop piece was generated by *bulge-skip building steps* to permit the incorporation of bulged nucleotides. For a 5′-loop piece $(i, k)$, models were created by building nucleotide $k$ with a bulge at $k–1$ from the models of the smaller loop piece $(i, k–2)$. Similarly, for a 3′-loop piece $(l, j)$, models were built from the models of the smaller loop piece $(l+2, j)$. The recursion proceed until the loop is completed [i.e, the full length loop $(i, j)$].

iii.    Additional models of the full-length loop were generated by combining models of loop piece $(i, k–1)$ with models for loop piece $(k+1, j)$ and applying the *single-nucleotide building step* to build nucleotide $k$ from either the 5′ edge $(k-1)$ or the 3′ edge $(k+1)$.

iv.     For each sub-region, all available models were grouped in a *clustering step*, with the lowest 1000 energy models carried forward.

v.      All models of the entire loop were grouped in a final *clustering step*.

The initial conditions for the recursion were defined by the crystallographic model with the entire loop ($i, j$) excised, as well as the P, O1P, O2P, O5´ atoms of neighboring nucleotide $j+1$ removed (to erase information about the "take-off" direction for the loop). This initial model also contained no hydrogens, as these atoms are not typically included in crystallographic models; 2´-OH hydrogens for the entire structure were sampled in the building steps (see below). The individual sampling steps are described next.

*(b) Single-nucleotide building step*

Here, the nucleotide to be built was directly adjacent to the 5´ or 3´ edge of the previous sub-region. Sub-Angstrom enumeration of the nucleotide's conformational space was achieved by sampling all the backbone torsions connecting the nucleotide to its 5´ or 3´ neighbor [the backbone "suite" $\varepsilon, \zeta, \alpha, \beta, \gamma$ (1)], the nucleobase's glycosidic torsion ($\chi$) in 20° intervals, and the nucleotide ribose sugar (torsions $\nu_0, \nu_1, \nu_2, \nu_3/\delta,$ and $\nu_4$) in its north (3´-endo) and south (2´-endo) puckered conformations. 2´-OH torsion sampling was carried out separately (see below). The $\alpha, \gamma,$ and $\zeta$ torsions were allowed to sample the full 360° range. Other torsions were restricted to sterically allowed regions: $\beta$ in the *trans* region (80°–280°) and $\varepsilon$ to the *trans* and *gauche$^+$* regions (170°–290° if the sugar pucker was north; 182°–302° if south). For both pyrimidines and purines, the $\chi$ torsion was sampled in the *anti* region (59°-99° if north; 97°-137° and south). Additionally, for purines, the $\chi$ torsion was sampled in the *syn* region (289°-329° if north; 290°-330° if south). This led to the generation of 5,388,768 unique conformations for purines and 2,694,384 unique conformations for pyrimidines that were spaced from each other, on average, by 0.6 Å all-heavy-atom RMSD (Table S5).

After the enumeration above, every atom in the nucleotide was positioned except for the single 2´-OH hydrogen atom. The Rosetta *packer* algorithm (2, 3) originally written for protein side-chain packing and design, was used determine the optimal orientation of these atoms in all loop nucleotides as well as in all surrounding nucleotides through efficient combinatorial sampling, similar to REDUCE (4). The resulting models were then finely clustered with an all-atom RMSD cutoff of 0.5 Å over all the sampled atoms and by similarity of the newly built nucleotide's sugar puckers and the lowest energy members of each group were retained. The 108 lowest energy conformations (typically including all conformations within 15 $k_B T$ of the very lowest energy model in the calculation) were subject to continuous minimization over all loop torsions with the Davidson–Fletcher–Powell algorithm (the Rosetta *minimizer*).

The calculations above, for even a single nucleotide, require comparing energies of millions of conformations in order to select out the lowest energy models. To accelerate the computation, we took advantage of the working hypothesis that the newly built nucleotides should form at least one favorable and no unfavorable interactions with previous nucleotides. We imposed the following filters. First, conformations were discarded in which the new nucleotide could not potentially form either base-stacking or base-pairing interactions with previously built nucleotides, as assessed by geometric criteria that we defined based on interactions observed in the RNA crystallographic database (see also (5)). Define $d$ as the distance between base centroids; $z_1$ and $z_2$ are the distance vector components projected onto the first base normal and second base normal; $\rho_1$ and $\rho_1$ are the distance vector components radial to the base normal; and $\theta$ is the angle between the base normals. The base-stacking criteria were defined to loosely include geometries for co-axially stacked bases: $d < 6.4$ Å; $2.5$ Å $< |z_1| < 4.5$ Å and $2.5$ Å $< |z_2| < 4.5$ Å; $|z_1|/d > 0.707$ and $|z_2|/d > 0.707$; and cos $\theta > 0.707$. The base-pairing criteria were $d < 12.0$ Å; $\rho_1 < 5.0$ Å and $\rho_2 < 5.0$ Å; $|z_1| < 5.0$ Å and $|z_2| < 5.0$ Å; $|z_1|/d < 0.5$ and $|z_2|/d < 0.5$; and cos $\theta > 0.866$. As a second filter, nucleotide conformations were discarded in which the attractive component *fa_atr* of the computed van der Waals interaction of the new nucleotide with previous nucleotides was not better than $-1.0$ Rosetta units ($\sim 1$ $k_BT$) or if the sum of the attractive and repulsive component *fa_rep* was worse than $0.0$ Rosetta units. Approximately 1% of all the conformations passed the filters; computing the full Rosetta all-atom energy score for the resulting tens of thousands of conformations required only tens of minutes on an Intel Core i7 2.66 GHz processor. Test calculations (on single-nucleotide building steps) run without the filters gave indistinguishable results at much greater computational expense.

*(c) Bulge-skip building step*

To permit modeling of single bulged residues, a bulge-skip building step was implemented, involving two nucleotides. This algorithm involved finding all possible positions of a new nucleotide that was covalently linked to previously built nucleotides by a single bulged (interaction-free) nucleotide. The new nucleotide was sampled by carrying out a grid search over the nucleobase's six rigid-body degrees of freedom. The translation component (*x, y,* and *z* coordinates of the base centroid) was sampled in 1.0 Å intervals over a grid encapsulating all previously built nucleotides; the rotation component (the three

Euler angles: azimuthal $\phi$, bend angle $\theta$, and second azimuthal $\psi$) was sampled over 20° intervals over the entire 0°–360° range for $\phi$ and $\psi$, and in intervals of 0.05 between –1.0 and 1.0 over cos $\theta$ in 0.05 intervals, to prevent oversampling of polar regions (6). For each of the resulting base positions/orientations, the nucleotide's ribose sugar was built by sampling the glycosidic torsion ($\chi$) in 20° intervals and by sampling the ribose sugar in both the north and south conformations. Exactly as above, the 2´-OH positions were sampled by the Rosetta *packer*; conformations were clustered; and the lowest energy 108 models were selected. All loop torsions (and the position and orientation of the newly built centroid) were then subjected to continuous minimization, and the conformations were re-clustered. A place-holder "bulge" nucleotide suite (the nucleotide plus the phosphate group of its 3´ neighbor) was built to connect the new sampled nucleotide and the previously built nucleotides. The bulge nucleotide suite was modeled as having an undefined position (all atoms of this nucleotide suite were assigned the Rosetta "virtual" atom type) and a fixed energy bonus was given to account for conformational entropy (see below). Again, a series of filters accelerated the calculation. Conformations were discarded in which the distance between the built nucleotide and the previous nucleotide would disallow building of the intervening bulge nucleotide and chain closure (distance between the C5´ atom of the 3´ loop nucleotide and the O3´ atom of the 5´ nucleotide beyond 11.4 Å). A further coarse filter discarded nucleotide conformations with more than 2 atoms that clashed with previously built atoms (distance less than sum of the atom van der Waal radii minus 0.8 Å). As with the single-nucleotide building step, conformations were discarded that did not satisfy either a similar base-stacking criteria as above (with the cutoffs on $|z|/d$ and cos $\theta$ increased to 0.90) or both the same base-stacking and base pairing criteria (with the cutoff on cos $\theta$ loosened to 0.707).

*(d) Chain closure, for building steps that result in completed loops.*

Chain closure was required at several points in the calculations. When the single-nucleotide or the bulge-skip building step corresponds to building the last remaining nucleotide in the loop (i.e. completing the loop), the chain needs to be closed. Both building steps were modified as follows to allow for chain closure. First, for the *chain-closure+single-nucleotide building step*, conformations that fail the filter for base-stacking or base-pairing interactions were not discarded to allow for the possibility that the new nucleotide is a bulge (see Supporting Methods). Second, for both the *chain-closure+single-nucleotide* *building step* and the *chain-closure+bulge-skip building step*, the van der Waals interaction filter was modified so that a nucleotide conformation was discarded only if the repulsive component *fa_rep* of the van der Waals interaction was worse than 10.0 Rosetta units. Chain closure was carried out via cyclic coordinate descent (CCD) (7) on the torsion angles of closing nucleotide's backbone suite ($\varepsilon$, $\zeta$, $\alpha$, $\beta$, $\gamma$). Two additional filters were applied to discard conformations that do not have the chain properly closed. First, before CCD chain closure was applied, a fast distance filter was used to discard conformations in which the distance between the C5′ atom of the nucleotide 3′ of the chain-break and O3′ atom of the nucleotide 5′ of the chain-break was greater than the theoretical maximum distance (4.63 Å) or less than the theoretical minimum distance (2.00 Å). After the attempted chain-closure with CCD, nucleotides with chain conformations that were not properly closed were discarded, i.e., if the O3′-P bond distance deviated greater than 0.15 Å from the ideal value (1.593 Å) or $(\Delta\theta_1/8.5°)^2 + (\Delta\theta_2/5.7°)^2 \geq 5.0$, where $\Delta\theta_1$ and $\Delta\theta_2$ are the deviations of the C3′-O3′-P and O3′-P-O5′ bond angles from their ideal values (119.8° and 108.97°, respectively). Subsequent minimization of loop torsion angles (see above) with the Rosetta *linear_chainbreak* energy term (see below) further improved the geometry at the closed nucleotide suite.

*(e) Clustering*

For each sub-region, all the generated conformations were clustered together with stringent criteria. Two conformations were grouped together if their all-heavy-atom RMSD over the whole loop motif was less than 0.7 Å, their all-heavy-atom RMSD computed over each individual nucleotide was less than 1.0 Å, and if the north/south sugar pucker classification matched over each nucleotide. The lowest energy members of each of the 1000 lowest energy clusters were then kept and used as starting structures for the next building step along the pathway. For the final *clustering* step, the conformations of all full length loop regions were combined and clustered together with a 1.5 Å RMSD over the whole loop motif and a 2.5 Å RMSD clustering over each individual nucleotides. The lowest energy members of each of five lowest energy clusters were designated as cluster representatives and predicted models.

Explicit command line examples for SWA loop modeling in Rosetta are given below [see subsection, *Generating SWA models (command lines)*].

**FARFAR**

FARFAR models were generated by fragment assembly followed by full-atom refinement in the Rosetta framework, as described previously (2); the fragment source was the large ribosomal subunit of *H. marismortuii* (PDB: 1JJ2), filtered to remove loops with evolutionary kinship to targets in our benchmark. 250,000 models were generated per motif, and clustered together with a 1.5 Å RMSD over the whole loop motif and 2.5 Å RMSD clustering over each individual nucleotides. The lowest energy members of each of five lowest energy clusters were designated as cluster representatives and predicted models. To ensure a rigorous comparison to the SWA results above, the same torsion angles were sampled ($\varepsilon$ and $\zeta$ of the nucleotide 5′ to the loop; all torsions inside the loop, which was built with Rosetta ideal bond lengths and bond angles; $\alpha$, $\beta$, and $\gamma$ of the nucleotide 3′ to the loop; and all 2′-OH torsions). Fragment assembly of loops requires a transient chainbreak that is iteratively closed after each move (8, 9); as in SWA, we chose cutpoint locations at any of the possible loop suites with equal probability, and carried our CCD loop closure. Explicit command line examples for FARFAR loop modeling are below [see subsection, *Generating FARFAR models (command lines)*].

### Generation of optimized native models

To fairly compare the computed energies of crystallographic (native) models and *de novo* models, both models need to be optimized with the Rosetta all-atom energy function over the same degrees of freedom with similar amounts of computational power. We optimized the native loop structure using three different methods. First, 50,000 FARFAR models were generated using only fragment derived from the native loop structure. Second, 50,000 FARFAR models were generated using the standard fragment library doped with native fragments, and models within 1.5 Å all-heavy-atom RMSD to the crystallographic structure were kept. Third, SWA was carried out on the loops, but focused on conformations near the crystallographic models; a filter was imposed at each building step requiring models to be within 2 Å RMSD of the crystal structure. The lowest energy model derived from all three methods was taken as the optimized native conformation.

### Optimization of the Rosetta all-atom energy function

All optimization was carried out in the Rosetta all-atom energy function (2), with small improvements. The energy unit reported herein is in Rosetta units (RU), which is used internally by the Rosetta program to store evaluated energy values. Comparisons to RNA Watson-Crick helix thermodynamic parameters

(10) indicate that 1 Rosetta unit is approximately equal to 1 $k_B T$ (2) [the structure modeling results given in the main text do not depend on the absolute scale of this energy unit]. Compared to prior work (2), the internally used Rosetta weighting factor of two energy terms were modified:

a. *Torsional Potential term*: The Rosetta weighting factor of the torsional potential term *rna_torsion* was poorly constrained in the original optimization, with weighting factors ranging from 0.1 to 5 giving similar accuracies in FARFAR *de novo* modeling (2). The original weighting factor was chosen to be 0.1 and with this weighting factor, the torsional potential gives negligible energetic contribution; for example the energetic difference between the energy minima and the maxima for each torsional angle was on the order of 0.1 RU for most torsions. In this work, we increased the Rosetta weighting factor of this energy term from 0.1 to 2.9.

b. *Side-chain-side-chain and long-range backbone-side-chain hydrogen bond terms*: The Rosetta weighting factor for both of the terms *hbond_sc* and *hbond_bb_sc* were set to 2.4, instead of 3.4 in (2); this change in the weighting factor was found to slightly improve the results obtained in (2).

Additionally, three new energy terms were introduced:

a. *Linear chainbreak term*. This term was introduced to penalize conformations in which the chain does not properly close at the chain-closure location. Analogous to loop closure in Rosetta protein modeling (8, 9), chain-closure was assessed and optimized by computing *linear_chainbreak,* the summed distances of virtual atoms OVL1 and OVL2 appended to the 5′ nucleotide with the P and O5′ of the 3′ nucleotide, and of virtual atom OVU1 prepended to the 3′ nucleotide with O3′of the 5′ nucleotide.

b. *Base-stacking term*. Previously, the only term that energetically favored base-stacking conformations was the attractive component of the Vander Waals interaction. The term *fa_stack* was introduced to approximately model the energetic contribution of π-π dispersion interactions beyond attraction already captured in the Rosetta van der Waals term; its functional form was set to approximately reproduce quantum mechanical calculations on parallel benzene-benzene dimers (11). For two heavy atoms in two different nucleobase, define $d$ to be the inter-atom distance, $z$ to be the component of the inter-atom vector parallel to the first base normal, and $\cos\phi = z/d$. An attractive potential $g(\cos\phi) \times f(d)$ was applied, where $g(\cos\phi) = \cos^2\phi$ ensures parallel stacking of bases, and $f(d) = -0.025$ for $d < 4$ Å, interpolating by a standard cubic spline to $f(d) = 0.0$ for $d > 6$ Å.

c. *Bulge bonus term*. Conformational entropy is not explicitly calculated in the Rosetta energy function; in prior protein and nucleotide work, conformational entropies were assumed to be similar for all well-packed conformations. Nevertheless, RNA loops and noncanonical motifs often exhibit bulged bases, which retain conformational fluctuations compared to bases that are involved in base-pairing and/or base-stacking interactions. To account for this, bulge nucleotides were given a fixed bonus *rna_bulge* (here, −4.5 RU; similar results were achieved with −3.0 to −6.0 RU). In SWA, a nucleotide suite (the nucleotide plus the phosphate group of its 3′ neighbor) was assigned to be a bulge if (1) the nucleotide was the bulge in the *bulge-skip building step* or (2) if the nucleotide was built in the *single-nucleotide + chain-closure building step* and did not pass the filter for base-stacking or base-pairing interactions. In FARFAR, a nucleotide suite was assigned to be a bulge if the fixed-bulge bonus outweighed the total energetic contribution of the nucleotide suite (assessed by assigning the Rosetta "virtual" atom type to the nucleotide suite and recalculating the Rosetta energy of the structure). Due to their rarity in experimental structures, we did not model consecutive bulges in either SWA or FARFAR.

### Modeling peripheral regions

To optimize the computational speed of both the SWA and FARFAR algorithms, especially for cases taken from the large ribosomal subunit, we modeled peripheral regions well outside the loop motif only as steric (nucleotides in which every atoms were beyond distance $d$ from the loop; $d = 10$ Å for the rRNA cases and $d = 5$ Å for the other cases). We retained a 3D grid of these "steric-only" regions and screened out conformations in which any loop nucleotide contained more than 2 atom-atom pair clashes (distance less than sum of the atom van der Waal radii minus 1.2 Å) with any of these peripheral nucleotides. This optimization was applied to both SWA and FARFAR to ensure comparability between the two methods. We ran SWA on several test cases, including the J2/4 loop of the TPP riboswitch, with and without this optimization and found indistinguishable results. This optimization was not used in the C7.2 tetraloop receptor modeling runs.

### Extending the SWA method to treat hairpins and multiple-stranded loops

The Stepwise Assembly method was extended to treat hairpins (loop beginning and ending at the same helix) as well as multiple-stranded loops. Briefly, the overall motif building calculation was ordered into $N^2$ building stages corresponding to each continuous subregion of the target motif of length $N$ nucleotides.

The same single-nucleotide building, bulge-skip building, chain-closure, and the models clustering steps described above were implemented. The canonical base pairs at the edges of the motif were assumed to be known *a priori* (2) and are built using idealized geometry (12). An extensive benchmark testing the SWA method on hairpins and multiple-stranded loops is underway; initial results are presented in SI Fig. S5.

***Generating Stepwise Assembly models (command line scripts):***

Documentation of the stepwise assembly code is being made available with the Rosetta software. For completeness, we provide here explicit examples of command line scripts for modeling the three-nucleotides loop in the C7.2 tetraloop receptor. The same scripts were used in modeling loop motifs in the benchmark.

a. The entire SWA loop building process can be formulated as a directed acyclic graph (DAG) (13) with the command files for each individual building step automatically set up with the Python script `setup_SWA_RNA_dag_job_files.py`. For the C7.2 tetraloop receptor case, the setup script command line is:

```
setup_SWA_RNA_dag_job_files.py -s template.pdb —fasta C7_2_target.fasta -sample_res
11 12 13 -nstruct 1000 -num_slave_nodes 250 -single_stranded_loop_mode True
```

The "`-s`" flag specifies the template structure. In this case, the known crystallographic model of the P4-P6 *Tetrahymena* group 1 intron (PDB: 2R8S) was used as the template. Specifically, coordinates of nucleotides in the L5b tetraloop and J6a/J6b receptor regions of 2R8S were copied into `template.pdb` [nts 149-154, 221-224, 227-228 and 246-252 in conventional P4-P6 numbering (see Fig. S3C)]. The AA-platform (nts 225 and 226) was omitted since it will be replaced by the three-nucleotide GUA loop in the C7.2 model. We then mutate the G227-U247 base pair into a C-G base pair (to match C7.2 sequence) by performing base mutations inside Rosetta (preserving the glycosidic torsion and the sugar-phosphate backbone). The "`—fasta`" flag specifies the fasta file containing the full sequence of the target structure:

```
> C7_2_target.fasta
ggaaacuccuguacuagaugga
1   5   10   15   20
```

The "`-sample_res`" flag specifies the missing nucleotides in the template structure that will be modeled

*de novo*; the nucleotides "`11 12 13`" corresponds to the three-nucleotide GUA loop. The "`-nstruct`" flag specifies the number of cluster centers to be carried forward, the "`-num_slave_nodes`" flag specifies the number of CPUs to be allocated to the job, and the "`-single_stranded_loop_mode`" flag specifies that the motif is a single-stranded loop.

Once the DAG files are set up, the Python script `dagman_continuous.py` is then called. This script controls the workflow of the DAG and automatically queues the individual building steps on a high-performance Linux cluster using either the Load Sharing Facility or Condor queuing systems. Examples of the automatically generated command lines for each individual step are provided below.

b. Example of an automatically generated command line for the single-nucleotide building step (building G11 from the 5′ end):

```
rna_swa_test.<exe> -algorithm rna_sample -database <path to database> -s template.pdb
-out:file:silent REGION_0_1/START_FROM_REGION_0_0/region_0_1_sample.out -
cluster:radius 0.5 -score:weights rna_loop_hires_04092010.wts -fixed_res 1 2 3 4 5 6
7 8 9 10 14 15 16 17 18 19 20 21 22 -rmsd_res 11 12 13 -jump_point_pairs 1-22 -
alignment_res 1-22 —fasta C7_2_target.fasta -global_sample_res_list 11 12 13 -
sample_res 11 -input_res 1 2 3 4 5 6 7 8 9 10 14 15 16 17 18 19 20 21 22
```

c. Example of an automatically generated command line for the bulge-skip building step (building U12 with G11 as a bulge from the 5′ end):

```
rna_swa_test.<exe> -algorithm rna_sample -database <path to database> -s template.pdb
-out:file:silent REGION_0_2/START_FROM_REGION_0_0/region_0_2_sample.out -
cluster:radius 0.5 -score:weights rna_loop_hires_04092010.wts -fixed_res 1 2 3 4 5 6
7 8 9 10 14 15 16 17 18 19 20 21 22 -rmsd_res 11 12 13 -jump_point_pairs 1-22 -
alignment_res 1-22 -fasta C7_2_target.fasta -global_sample_res_list 11 12 13 -
sample_res 12 11 -floating_base true -input_res 1 2 3 4 5 6 7 8 9 10 14 15 16 17 18
19 20 21 22
```

d. Example of an automatically generated command line for the clustering step (after building G11 from the 5′ end):

```
rna_swa_test.<exe> -algorithm rna_cluster -database <path to database> -nstruct 1000
-clusterer_min_struct 1000 -suite_cluster_radius 1.0 -loop_cluster_radius 0.7 -
clusterer_quick_alignment true  -score:weights rna_loop_hires_04092010.wts -fixed_res
1 2 3 4 5 6 7 8 9 10 14 15 16 17 18 19 20 21 22 -rmsd_res 11 12 13 -jump_point_pairs
1-22 -alignment_res 1-22 -fasta C7_2_target.fasta -sample_res 11 -input_res 1 2 3 4 5
6 7 8 9 10 14 15 16 17 18 19 20 21 22 -in:file:silent
REGION_0_1/start_from_region_0_0_sample_filtered.out -in:file:silent_struct_type
binary_rna -out:file:silent region_0_1_sample.cluster.out -silent_read_through_errors
```

***Generating FARFAR models (command lines):***

a. The entire FARFAR loop building process can also be formulated as a directed acyclic graph (DAG), albeit a simple one since the FARFAR method creates models in an 'embarrassingly parallel' fashion with only one building step needed to complete the whole loop. The python script `setup_FARFAR_RNA_dag_job_files.py` is used to set up the dag files. For the C7.2 tetraloop receptor case, the setup script command line is:

```
setup_FARFAR_RNA_dag_job_files.py -s template.pdb —fasta C7_2_target.fasta -
sample_res 11 12 13 -nstruct 250000 -num_slave_nodes 250 -single_stranded_loop_mode
True
```

The "`-nstruct`" flag differs from the SWA case above in that it instead specifies here, the total number of models to be generated by FARFAR in an embarrassingly parallel fashion. Once the DAG files are set up, the Python script `dagman_continuous.py` is then called. This script controls the workflow of the DAG and automatically queues the individual building steps on a high-performance Linux cluster using either the Load Sharing Facility or Condor queuing systems.

b. Example of an automatically generated fragment assembly command line to build the entire loop:

```
rna_denovo.<exe> -database <path to database> -fasta C7_2_target.fasta -params_file
cutpoint_closed_13/params -nstruct 200 -cycles 10000 -output_virtual -heat true -
close_loops true -minimize_rna true -out:file:silent
cutpoint_closed_13/DAG_ID_0/silent_file.out -score:weights
rna_loop_hires_04092010.wts -in:file:silent_struct_type binary_rna -fixed_res 1 2 3 4
5 6 7 8 9 10 14 15 16 17 18 19 20 21 22 -virtual_phosphate_list 1 -allow_bulge_mode
true -allow_bulge_res_list 11 12 13 -rmsd_res 11 12 13 -native_alignment_res 1 22 -s
cutpoint_closed_13/FARFAR_start_cutpoint_13.out -start_silent_tag
FARFAR_start_cutpoint_13
```

The file `cutpoint_closed_13/params` contains information regarding which nucleotides are sampled and the loop cutpoint location:

```
OBLIGATE   PAIR 1 22 H H A
ALLOW_INSERT  11 13
CUTPOINT_CLOSED 13
```

In this case, the loop cutpoint location is at the phosphate backbone (O3′-P bond) between nucleotides A13 and C14. The possible `CUTPOINT_CLOSED` values for this 3-nucleotide loop are 10, 11, 12 and 13.

*Additional modeling the C7.2 tetraloop receptor with SWA and FARFAR*

We first modeled the three-nucleotide GUA loop in the C7.2 tetraloop receptor *de novo* using the standard SWA and FARFAR loop-modeling procedure as described above. While the standard SWA run let to satisfactory loop conformations, the generated models contain an incorrect geometry in the C227-G247 closing base pair (in conventional P4-P6 numbering). This incorrect geometry arises because in the standard loop-modeling procedure, we mutated the G227-U247 base pair into a C227-G247 base pair (to match C7.2 sequence) even through G/U wobble and Watson-Crick base pairs are not isosteric (14). Furthermore, in the standard loop-modeling procedure, the generated models inherit the relative position and orientation between the P6a and the P6b helix (see Fig. 4C) from the P4-P6 crystallographic model (PDB: 2R8S); however, the correct relative position and orientation between the two helices might be different for the C7.2 mutant. To resolves these issues, we carried out additional runs as follows. First, we replaced the two crystallographic base pairs G227-U247 and U228-A246 in the P6b helix with an idealized two base pairs C-G/U-A helix. Second, we sampled the relative position and orientation between the P6a and the P6b helix. For SWA, this is accomplished by including building steps that sampled the A6-C7 (see Fig. 4A) loop backbone suite and closed the chain with CCD (7) at the G10-A11 inter-helix backbone suite, and vice versa. This modified run gave similar loop conformations to the standard SWA loop-modeling run. For FARFAR, the G4-U5-A6 loop (including the U3-G4 and A6-C7 backbone suite) was sampled with fragment assembly and the chain was closed with CCD at the G10-A11 inter-helix backbone suite. The SWA and FARFAR models presented in main text Fig. 4B and SI Fig. S2 were generated using this modified procedure.

*Generating RLooM models:*

The website (http://rloom.mpimp-golm.mpg.de/) provides a web interface for homology RNA loop modeling with RLooM (database version 12-19-08) (15, 16). The modeling procedure involves two steps.

Step 1: Upload the template PDB to be used for modeling onto the web-server. For modeling the three-nucleotide loop in the C7.2 tetraloop receptor, the known crystallographic model of the P4-P6 *Tetrahymena* group 1 intron (PDB: 2R8S) was used as the template PDB. The AA-platform (nucleotides 225 and 226 in the conventional P4-P6 numbering) was removed from the template PDB and replaced with the three-nucleotide G-U-A loop in the extended conformation. For modeling the loop motifs in the benchmark, the template PDB is a parsed out segment of the crystallographic model containing the native loop and its surroundings.

Step 2: Input the command line. For example, the command line for modeling the three-nucleotide loop in the C7.2 tetraloop receptor is:

```

<anchor>122 :R</anchor>
<anchor>126 :R</anchor>
<query k=Δ_H>GUA</query>

```

Nucleotides 122 and 126 (224 and 227 in the conventional P4-P6 numbering) are the 5′ and 3′ anchor respectively. GUA is the sequence of the target loop. $\Delta_H$ is the maximum allowed sequence dissimilarity as given by the number of differing bases between the homolog loops found by RLooM and the target loop sequence (we started the search at $\Delta_H=0$ and increase this parameter until at least a total of five models were generated).

The default value for the maximum anchor-atom-RMSD (4.00 Å) and clash thresholds (5.00 Å) were used. By default, RLooM does not mutate the sequence of the generated loops to match the sequence of the target loop, so we performed this base mutation step inside Rosetta (preserving the glycosidic torsion and the sugar-phosphate backbone). The RLooM models were ranked first by sequence similarity to the target loop (from high to low) and second by anchor-atom-RMSD (from low to high). The top five ranked models were then taken as the predicted models. For modeling the loop motifs in the benchmark, we included every structure in the PDB as possible homology candidates except for directly related loop structures from the same species as the target loop, i.e., the homolog loop came from the same motif of the same biomolecule of the same species as the native loop. The "0.5 Å with sequence identity" cluster set was used, following ref (15).

***Generating ModeRNA models***

The ModeRNA (17) modeling python module (version 1.6.0) was downloaded from http://genesilico.pl/moderna/. ModeRNA models were then generated following the instructions located at http://genesilico.pl/moderna/tutorial. Inside the python interpreter:

```
>from moderna import *
>t = load_template('template.pdb', 'chain_ID')
>a = load_alignment('alignment.fasta') m = create_model(t,a)
>m.write_pdb_file('ModeRNA_model.pdb')
```

ModeRNA was used to model the three-nucleotide G4-U5-A6 loop (see Fig. 4A) in the C7.2 tetraloop receptor. The template structure was the known crystallographic model of the P4-P6 *Tetrahymena* group 1 intron (PDB: 2R8S), with the AA-platform (nucleotides 225 and 226 in the conventional P4-P6 numbering) removed. The inputted `chain_ID` was 'R' and the following alignment sequence was used:

```
>model sequence (C7.2 mutant):
GGAAUUGCGGGAAAGGGGUCAACAGCCGUUCAGUACCAAGUCUCAGGGGAAACUUUGAGAUGGCCUUGCAAAGGGUAUGGUAAUA
AGCUGACGGACAUGGUCCUAACACGCAGCCAAGUCCUGUACUCAACAGAUCUUCUGUUGAGAUGGAUGCAGUUCA

>template (PDB:2R8S with AA-platform removed):
GGAAUUGCGGGAAAGGGGUCAACAGCCGUUCAGUACCAAGUCUCAGGGGAAACUUUGAGAUGGCCUUGCAAAGGGUAUGGUAAUA
AGCUGACGGACAUGGUCCUAACACGCAGCCAAGUCCU---GUCAACAGAUCUUCUGUUGAUAUGGAUGCAGUUCA
```

Fig. S2F shows the ModeRNA model for the C7.2 mutant at the remodeled tetraloop receptor region.

***Generating atom-atom pair clash list with Molprobity***

The *clashlist* shell script (downloaded from http://kinemage.biochem.duke.edu/software/scripts.php, also implemented as part of the Molprobity program (18, 19)), was used to identify clashed atom-atom pairs in the C7.2 tetraloop receptor models generated by SWA, FARFAR, RLooM and ModeRNA. First, the original hydrogen atoms in the PDB file (if any existed) were removed and the program REDUCE (4) was then used to add hydrogen atoms back into the models. Second, Molprobity was used to revise the PDB to version 2.3 standards. Third, the *clashlist* shell script was run with the optional "–stdbonds" flag [*clashlist* does not identify severe steric clashes, where the two atom-centers are very close together, if this flag is omitted (19)].

***Generating single nucleotide conformations for clusters counting***

The conformations used for the clusters counting exercise (see SI Table. S5) were generated in the following way. First, an anchor adenosine nucleotide was built in the A-form conformation. A 'moving' adenosine nucleotide was then attached to the 3′ end of the anchor nucleotide. All backbone torsions between the anchor and moving nucleotide ($\alpha$, $\beta$, $\gamma$, $\varepsilon$, $\zeta$) and the glycosidic torsion ($\chi$) of the moving nucleotide were sampled in 20° intervals using the sampling procedure of the single-nucleotide building step (see *Stepwise Assembly in Rosetta* section), leading to the generation of 5,388,768 conformations. The 2′-OH torsion was neglected since it only determines the position of the 2′-OH hydrogen, a non-heavy atom. These conformations were then filtered for steric clashes both intra-nucleotide and between the moving nucleotide and the A-form anchor (a conformation was discarded if four or more atom-pairs have their van der Waal radii overlap by more than 0.5 Å) leaving a total 2,183,676 free-clash conformations, which were then clustered.

***P4-P6 sequence used in the chemical accessibility experiments:***

1.  Wild type P4-P6 domain of *Tetrahymena* Group I Intron:

```
5'-ggccaaaacaacGGAAUUGCGGGAAAGGGGUCAACAGCCGUUCAGUACCAAGUCUCAGGGGAAACUUUGAGAUGGCCUUGCA
AAGGGUAUGGUAAUAAGCUGACGGACAUGGUCCUAACCACGCAGCCAAGUCCUAAGUCAACAGAUCUUCUGUUGAUAUGGAUGCA
GUUCAaaaccaaacaaagaaacaacaacaacaac-3'
```

2. C7.2 mutant:

```
5'-ggccaaaacaacGGAAUUGCGGGAAAGGGGUCAACAGCCGUUCAGUACCAAGUCUCAGGGGAAACUUUGAGAUGGCCUUGCA
AAGGGUAUGGUAAUAAGCUGACGGACAUGGUCCUAACCACGCAGCCAAGUCCUGUACUCAACAGAUCUUCUGUUGAGAUGGAUGC
AGuucaaaaccaaaccaaagaaacaacaacaacaac
```

The upper-case regions correspond to the P4-P6 domain, and flanking sequences designed to avoid base pairing interactions are shown in lowercase. The primer binding site is underlined.

# Supporting Results

### *Analysis of RLooM Models*
RLooM successfully modeled 2 of the 15 loop motifs in the benchmark (SI Table S2). RLooM modeled these 2 success cases by taking advantage of the similarity between the native loops (from 23s rRNA of *H. marismortui*) and homologous loops (at corresponding positions in 23s rRNA of *E. coli* and *D. radiodurans*). However, the failure to recover the native loop structure in the majority of the cases (13 of 15) suggested that homology RNA loop modeling is not a generally applicable strategy (given the limited number of RNA structures currently in the PDB). Some RNA structures in the benchmark including the SAM-II riboswitch and the M-box riboswitch currently do not contain homologous structures in the PDB (excluding directly related structures from the same species; last checked on April 27, 2011). Furthermore, even in cases where homologous structures exist in the PDB, the native and corresponding homolog loops might differ significantly sequence-wise and/or structure-wise. The corresponding homolog loop might also be completely missing in the homologous structures. For example, the J5/5a hinge motif in the benchmark was taken from the structure of the *Tetrahymena* Group I Intron. Homologous *Twort* and *Azoarcus* Group I Intron structures exist in the PDB, however, the J5/Ja hinge motif does not exist in these homologs. Another example is the J2/4 loop in the *A. thaliana* TPP riboswitch structure. The homologous *E. coli* TPP riboswitch structure contains a loop at the corresponding position, however significant sequence differences (UUGAA vs. UAUCA) prevent this homolog loop from being correctly identified. Lastly, for three of the five 23S rRNA loops in the benchmark (loop nucleotides 531-536, 1976-1985 and 2534-2540), RLooM fails to detect the correct homolog loops in the PDB. While correct homolog loops do in fact exist in the RLooM database, they are defined as sub-regions of larger loops in the database.

For modeling C7.2 tetraloop receptor, the top five RLooM models were generated. However, all the models were either invalidated by the chemical accessibility data and/or were poor in energy due to significant steric clashes and an energetically unfavorable empty cavity in the core of the structure. Fig. S2E shows the RLooM top ranked model for the C7.2 mutant at the remodeled tetraloop receptor region. Molprobity identified 63 atom-atom pair clashes in this model (excluding those inherited from the crystallographic model).

In modeling the three-nucleotide loop, the 5′ anchor (U3) does not form a Watson-Crick or G/U wobble base pair and this might affect RLooM's performance (for rationale, see (15)). We therefore also modeled the 4-nucleotide U3-G4-U5-A6 loop (see nucleotide numberings in Fig. 4A). However, the top 5 RLooM models generated by this alternative method also contained severe steric clashes and were not consistent with the chemical accessibility data.

**Table S1.** Loop Motifs Benchmark Information.

| Motif Name | Source | | # Nucleotides | | | # Base Pairs[a] | | # Potential Hydrogen Bonds[b] | | | | |
|---|---|---|---|---|---|---|---|---|---|---|---|---|
| | PDB:Chain (Nucleotide Segment) | Crystal Res[c] | Total | Bulge | Outlier[d] | NWC | WC or GU | Total | Base-Base | Base-2′OH | Base-Phos[e] | Others[f] |
| 5′ J1/2, Leadzyme | 1NUJ:E (23-26) | 1.80 Å | 4 | 0 | 0 | 4 | 0 | 14 | 8 | 2 | 4 | 0 |
| 5′ P1, M-Box Riboswitch | 2QBZ:X (164-167) | 2.60 Å | 4 | 1 | 3 | 4 | 0 | 11 | 5 | 3 | 1 | 2 |
| 3′ J5/5a, Group I Intron | 2R8S:R (196-199) | 1.95 Å | 4 | 1 | 1 | 4 | 0 | 10 | 6 | 2 | 0 | 2 |
| 5′ J5/5a, Group I Intron | 2R8S:R (122-126) | 1.95 Å | 5 | 1 | 3 | 4 | 0 | 11 | 7 | 2 | 0 | 2 |
| Hepatitis C Virus IRES IIa | 2PN4:A (53-57) | 2.32 Å | 5 | 0 | 1 | 0 | 0 | 4 | 0 | 4 | 0 | 0 |
| J2/4, TPP Riboswitch | 3D2V:B (40-44) | 2.00 Å | 5 | 1 | 2 | 4 | 0 | 17 | 7 | 3 | 5 | 2 |
| 23S rRNA (44-49) | 1S72:0 (44-49) | 2.40 Å | 6 | 1 | 0 | 5 | 0 | 21 | 10 | 6 | 5 | 0 |
| 23S rRNA (531-536) | 1S72:0 (531-536) | 2.40 Å | 6 | 0 | 1 | 4 | 0 | 16 | 5 | 3 | 5 | 3 |
| J3/1, Glycine Riboswitch | 3OWI:B (75-81) | 2.85 Å | 7 | 0 | 0 | 3 | 0 | 7 | 3 | 2 | 2 | 0 |
| J2/3, Group II Intron | 3G78:A (287-293) | 2.80 Å | 7 | 0 | 3 | 6 | 0 | 20 | 10 | 5 | 1 | 4 |
| L1, SAM-II Riboswitch | 2QWY:B (8-14) | 2.80 Å | 7 | 0 | 2 | 8 | 0 | 18 | 11 | 4 | 0 | 3 |
| L2, Viral RNA Pseudoknot | 1L2X:A (19-25) | 1.25 Å | 7 | 1 | 2 | 5 | 0 | 15 | 6 | 6 | 2 | 1 |
| 23S rRNA (2534-2540) | 1S72:0 (2534-2540) | 2.40 Å | 7 | 0 | 2 | 7 | 0 | 19 | 9 | 2 | 5 | 3 |
| 23S rRNA (1976-1985) | 1S72:0 (1976-1985) | 2.40 Å | 10 | 0 | 3 | 6 | 0 | 25 | 8 | 6 | 7 | 4 |
| 23S rRNA (2003-2012) | 1S72:0 (2003-2012) | 2.40 Å | 10 | 2 | 5 | 6 | 2 | 37 | 14 | 8 | 11 | 4 |
| Total | | | 94 | 8 | 28 | 70 | 2 | 245 | 109 | 58 | 48 | 30 |
| Per nucleotide | | | 1.00 | 0.09 | 0.30 | 0.74 | 0.02 | 2.61 | 1.16 | 0.62 | 0.51 | 0.32 |

[a] Base pairs are annotated using an automated method based the scheme of Leontis and Westhof (RNA. 2001 Apr;7(4):499-512.). The non-Watson-Crick base pair column ("NWC") excludes canonical A-U and G-C base pairs as well as G-U wobble base pairs.

[b] An automated method is used to identify polar hydrogen bonds with the cutoff following criteria: (a) the distance between the hydrogen atom and acceptor atom is less than 2.7 Å and (b) the donor-hydrogen-acceptor angle is greater than 130°.

[c] Resolution of the x-ray diffraction crystallographic data.

[d] An outlier is a nucleotide suite that is not part of the 46 most commonly observed RNA rotamers defined by the RNA Ontology Consortium (RNA. 2008 Mar;14(3):465-81)

[e] Base-Phos: Polar hydrogen bonds between the base and the phosphate group.

[f] Others: Polar hydrogen bonds between (a) Base and O4′ atom, (b) 2′-OH group and 2′-OH group, (c) 2′-OH group and phosphate group or (d) 2′-OH group and O4′ atom.

**Table S2**. Application of RLooM homology modeling to the loop motifs benchmark.

| Motif name | Motif properties | | | Top five RLooM models[a] | Excluded native source PDB[c] |
|---|---|---|---|---|---|
| | Length | PDB | Lowest RMSD[b] (Å) | Models information (template PDB, sequence dissimilarity, anchor-atoms-RMSD (Å)) | |
| 5´ J1/2, Leadzyme | 4 | 1NUJ | 4.45 | (1XJR, 0, 0.94), (1LDZ, 0, 0.94), (1VOW, 0, 1.26), (2B9N, 0, 1.60), (1SML, 0, 0.85) | none |
| 5´ P1, M-Box Riboswitch | 4 | 2QBZ | 3.96 | (2GYC, 0, 1.27), (1VOW, 0, 1.48), (2GIO, 0, 1.91), (2HGH, 1, 1.29), (1NJN, 1, 1.42) | 2QBZ |
| 3´ J5/5a, Group I Intron | 4 | 2R8S | 6.97 | (3DLL, 0, 1.68), (2ZJP, 0, 1.69), (2VHN, 0, 2.72), (2GYC, 0, 3.20), (3CF5, 0, 1.72) | 2R8S, 1X8W, 1GID, 1GRZ, 1L8V, 1HR2 |
| 5´ J5/5a, Group I Intron | 5 | 2R8S | 5.44 | (1PNY, 0, 1.63), (2AW4, 1, 1.64), (1NJM, 0, 1.79), (1VOU, 1, 1.91), (2OGN, 1, 2.29) | 2R8S, 1HR2, 1GID, 1L8V, 1X8W, 1GRZ |
| Hepatitis C Virus IRES IIa | 5 | 2PN4 | 3.30 | (1P5P, 0, 0.65), (1NKW, 1, 2.09), (2O43, 1, 2.19), (2RKJ, 1, 2.38), (2HGJ, 1, 2.518) | none |
| J2/4, TPP Riboswitch | 5 | 3D2V | 2.64 | (2GYC, 0, 7.54), (2GYC, 1, 1.29), (1C2W, 1, 1.45), (1C2W, 1, 1.90), (2HGJ, 1, 3.49) | 3D2X, 2CKY |
| 23S rRNA (44-49) | 6 | 1S72 | **1.42** | (1JZX, 0, 0.18), (1P9X, 0, 0.23), (1NJP, 0, 0.26), (1Z58, 0, 0.28), (2OGN, 0, 0.29), | 1YHQ, 1Q86 |
| 23S rRNA (531-536) | 6 | 1S72 | 8.45 | (2FEY, 1, 5.85), (1VQ5, 1, 6.25), (1YL3, 2, 1.98), (2JL8, 2, 2.02), (2HGQ, 2, 2.04) | none |
| J3/1, Glycine Riboswitch | 7 | 3OWI | 6.76 | (2VHP, 1, 4.91), (1C2W, 1, 12.95), (2V47, 2, 2.34), (1YI2, 2, 2.35), (2J01, 2, 2.40) | none |
| J2/3, Group II Intron | 7 | 3G78 | 12.49 | (2QP0, 1, 5.24), (2VHO, 1, 5.25), (3BBN, 2, 2.77), (2GY9, 2, 2.93), (1N36, 2, 3.03) | none |
| L1, SAM-II Riboswitch | 7 | 2QWY | 8.30 | (2VHM, 2, 2.29), (2GYC, 2, 2.95), (1C2W, 2, 7.68), (2VHM, 2, 9.40), (2QBE, 2, 9.48) | none |
| L2, Viral RNA Pseudoknot | 7 | 1L2X | 6.69 | (3EGZ, 2, 2.11), (3BBO, 2, 2.65), (1VSP, 2, 2.98), (1FKA, 2, 3.09), (3D5B, 2, 3.85) | none |
| 23S rRNA (2534-2540) | 7 | 1S72 | 10.19 | (2B64, 3, 1.99), (1JZX, 3, 2.06), (1YL4, 3, 2.30), (1KQS, 3, 2.43), (1P9X, 3, 2.65) | none |
| 23S rRNA (1976-1985) | 10 | 1S72 | 10.24 | (1VOR, 5, 2.78), (1VP0, 5, 2.97), (2V46, 5, 4.18), (1P9X, 5, 5.81), (3BBO, 5, 8.27) | none |
| 23S rRNA (2003-2012) | 10 | 1S72 | **0.87** | ( (2AW4, 2, 0.48), ( 2GYA, 2, 0.73), (2GYC, 2, 1.01), (1C2W, 2, 7.20), (1P9X, 3, 0.32) | 1M90 |
| **RMSD < 1.50 Å** | | | 2/15 | | |

[a] The top five models as ranked first by sequence similarity (from high to low) and second by anchor-atom-RMSD (from low to high). See SI Supporting Methods for explicit command-line examples used to generate the models. A complete analysis of the results is also given in SI Supporting Results.

[b] All-heavy-atom RMSD to the crystallographic loop. Nucleotides found to be bulged (both unpaired and unstacked) in the native crystallographic model were excluded from the RMSD calculation. Bold text indicates RMSD is within 1.5 Å of the crystal structure.

[c] We included every structure in the PDB as possible homology candidates except for structures presenting the same motif of the same biomolecule of the same species as the target native loop. The "0.5 Å with sequence identity" cluster set was used, following ref (Nucleic Acids Res. 2010 Jan;38(3):970-80).

**Table S3.** Supplemental loop motifs benchmark data (FARFAR).

| Motif Name | Motif properties | | Best of Five Lowest Energy Cluster Centers | | | | Lowest RMSD Model | | Lowest Energy Sampled | |
| --- | --- | --- | --- | --- | --- | --- | --- | --- | --- | --- |
| | Length | PDB | Cluster Rank | RMSD (Å) | Recovered BPs[a] | Rosetta Energy (RU) | RMSD (Å) | Recovered BPs[a] | Rosetta Energy (RU) | E-Gap to Optimized Native[b] (RU) |
| 5′ J1/2, Leadzyme | 4 | 1NUJ | 1 | 1.96 | 2/4 | -53.6 | 1.66 | 2/4 | -53.6 | 2.7 |
| 5′ P1, M-Box Riboswitch | 4 | 2QBZ | 3 | **0.72** | 4/4 | 129.0 | **0.53** | 4/4 | 128.9 | 2.3 |
| 3′ J5/5a, Group I Intron | 4 | 2R8S | 1 | **0.40** | 4/4 | -21.4 | **0.30** | 4/4 | -21.4 | **0.0** |
| 5′ J5/5a, Group I Intron | 5 | 2R8S | 2 | 4.08 | 2/4 | -51.3 | **1.05** | 3/4 | -52.5 | 0.3 |
| Hepatitis C Virus IRES IIa | 5 | 2PN4 | 3 | 2.11 | 0/0 | 27.3 | **1.04** | 0/0 | 25.4 | **-2.6** |
| J2/4, TPP Riboswitch | 5 | 3D2V | 1 | 6.66 | 2/4 | -83.4 | 1.74 | 2/4 | -83.4 | 10.8 |
| 23S rRNA (44-49) | 6 | 1S72 | 1 | **0.69** | 5/5 | -163.8 | **0.47** | 5/5 | -163.8 | 2.6 |
| 23S rRNA (531-536) | 6 | 1S72 | 1 | 3.18 | 2/4 | -253.8 | 2.44 | 1/4 | -253.8 | 6.9 |
| J3/1, Glycine Riboswitch | 7 | 3OWI | 1 | **1.13** | 3/3 | 94.8 | **0.71** | 3/3 | 94.8 | 2.5 |
| J2/3, Group II Intron | 7 | 3G78 | 1 | 1.72 | 5/6 | -59.0 | 1.53 | 4/6 | -59.0 | 8.5 |
| L1, SAM-II Riboswitch | 7 | 2QWY | 5 | 2.43 | 4/8 | 13.4 | **1.43** | 7/8 | 5.3 | 3.8 |
| L2, Viral RNA Pseudoknot | 7 | 1L2X | 4 | 5.44 | 2/4 | -183.4 | **1.35** | 3/5 | -185.6 | 3.7 |
| 23S rRNA (2534-2540) | 7 | 1S72 | 5 | 6.39 | 2/7 | -215.1 | 3.24 | 1/7 | -218.0 | 7.3 |
| 23S rRNA (1976-1985) | 10 | 1S72 | 1 | 11.54 | 0/5 | -283.5 | 5.08 | 0/5 | -283.5 | 13.0 |
| 23S rRNA (2003-2012) | 10 | 1S72 | 5 | 11.36 | 0/8 | -250.7 | 5.43 | 0/8 | -252.9 | 41.2 |
| AVERAGE | 6.3 | – | 2.3 | 3.99 | 2.5/4.7 | -90.3 | 1.87 | 2.6/4.7 | -91.5 | 6.9 |
| **RMSD < 1.50 Å** | – | – | – | 4/15 | – | – | 8/15 | – | – | – |
| **Energy Gap <= 0.0** | – | – | – | – | – | – | – | – | – | 2/15 |

[a] Number of native base-pairs correctly recovered by the *de novo* model. Base pairs are annotated using an automated method based the scheme of Leontis and Westhof (RNA. 2001 Apr;7(4):499-512.) and recovery entails having the correct edge-to-edge interaction (Watson-Crick, Hoogsteen, or Sugar Edge) and local strand orientation (*cis* or *trans*). Counts of correctly recovered base pairs are lowered owing to ambiguities in assigning bifurcated base pairs and pairs connected by single hydrogen bonds.

[b] The Rosetta energy of the lowest energy native loop structure optimized through three different methods (see SI Supporting Methods for details). Bold text indicates that the lowest energy sampled by the *de novo* run is lower than the energy of the optimized native (i.e. the energy gap is negative).

**Table S4**. Comparison of FARFAR with and without doping native fragments into the fragment library.

| Motif name | Motif properties | | Lowest RMSD model | |
|---|---|---|---|---|
| | Length | PDB | Exclude native fragments (standard) | Dope in native fragments (cheat[a]) |
| | | | RMSD (Å) | RMSD (Å) |
| 5´ J1/2, Leadzyme | 4 | 1NUJ | 1.66 | **0.30** |
| 5´ P1, M-Box Riboswitch | 4 | 2QBZ | **0.53** | **0.37** |
| 3´ J5/5a, Group I Intron | 4 | 2R8S | **0.30** | **0.30** |
| 5´ J5/5a, Group I Intron | 5 | 2R8S | **1.05** | **0.42** |
| Hepatitis C Virus IRES IIa | 5 | 2PN4 | **1.04** | **0.98** |
| J2/4, TPP Riboswitch | 5 | 3D2V | 1.74 | **0.45** |
| 23S rRNA (44-49)[A] | 6 | 1S72 | **0.47** | **0.38** |
| 23S rRNA (531-536)[A] | 6 | 1S72 | 2.44 | **0.46** |
| J3/1, Glycine Riboswitch | 7 | 3OWI | **0.71** | **0.49** |
| J2/3, Group II Intron | 7 | 3G78 | 1.53 | **0.95** |
| L1, SAM-II Riboswitch | 7 | 2QWY | **1.43** | **0.76** |
| L2, Viral RNA Pseudoknot | 7 | 1L2X | **1.35** | **0.43** |
| 23S rRNA (2534-2540)[b] | 7 | 1S72 | 3.24 | **0.49** |
| 23S rRNA (1976-1985)[b] | 10 | 1S72 | 5.08 | **0.69** |
| 23S rRNA (2003-2012)[b] | 10 | 1S72 | 5.43 | **0.48** |
| **RMSD < 1.50 Å** | | | 8/15 | 15/15 |

[a] Fragments of the native loop were doped/added into FARFAR's standard fragment library.

[b] FARFAR's standard fragment library is composed of RNA fragments extracted from a single structure of the archaeal large ribosomal subunit (PDB: 1JJ2). To mimic a true *denovo* modeling scenario, we ensure that regions with evolutionary kinship to our benchmark motifs were either absent or removed from the fragment database. For example, 5 motifs in the benchmark came from the PDB: 1S72 which is another archaeal large ribosomal subunit (in fact 1S72 is a revised structure of 1JJ2). Hence when modeling these 5 ribosomal loops, the correspond loop region in the PDB: 1JJ2 were excised from the fragment library.

**Table S5**. Number of conformation cluster centers in a nucleotide versus the RMSD cluster radius.

| All-atom RMSD cluster radius (Å) | # clash-free cluster centers |
|---|---|
| 4.0 | 69 |
| 3.5 | 129 |
| 3.0 | 294 |
| 2.5 | 803 |
| 2.0 | 2802 |
| 1.5 | 13784 |
| 1.0 | 124483 |
| 0.9 | 219,177 |
| 0.8 | 416,801 |
| 0.7 | 863,743 |
| 0.6 | 2,003,118 |
| 0.5 | 5,417,396 |
| 0.4 | 18,307,070 |
| 0.3 | 87,982,647 |
| 0.2 | 804,098,805 |
| 0.1 | 35,318,316,979 |

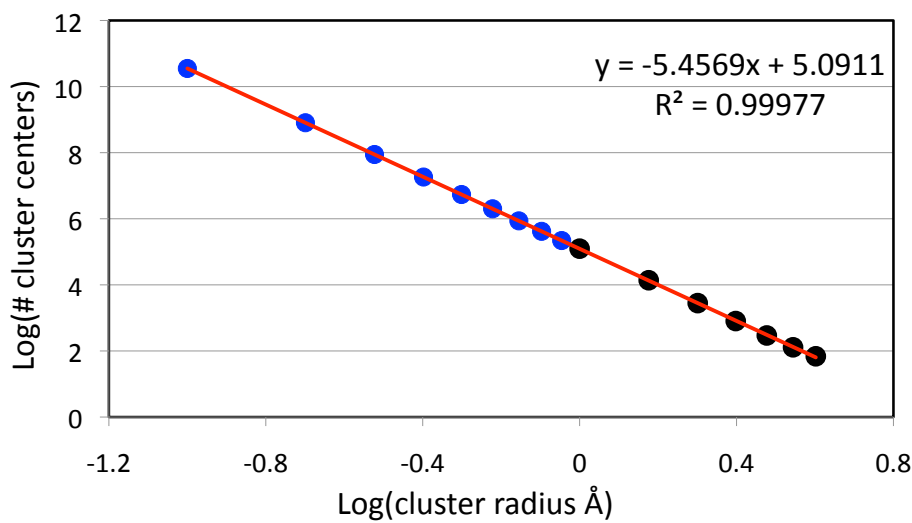

All sterically available conformations of a single RNA nucleotide were clustered together and the table reports the number of cluster centers as a function of the all-heavy-atom RMSD clustering radius. Clustering with sub-Angstrom threshold – as is necessary for high resolution modeling – leads to millions of unique clusters of single-nucleotide conformations. Data with cluster radius greater than or equal to 1.0 Å were directly generated (black points in plot). However, for smaller cluster radius values, the computation became infeasible and hence was determined through a least-squares interpolation (blue points in plot). Complete description of how the conformations used for this clusters counting exercise were generated is provided in the SI Supporting Methods: *Generating single nucleotide conformations for clusters counting* subsection.

**Table S6.** Supplemental loop motifs benchmark data (SWA).

| Motif Name | Motif properties | | Best of Five Lowest Energy Cluster Centers | | | | Lowest RMSD Model | | Lowest Energy Sampled | |
|---|---|---|---|---|---|---|---|---|---|---|
| | Length | PDB | Cluster Rank | RMSD (Å) | Recovered BPs[a] | Rosetta Energy (RU) | RMSD (Å) | Recovered BPs[a] | Rosetta Energy (RU) | E-Gap to Optimized Native[b] (RU) |
| 5′ J1/2, Leadzyme | 4 | 1NUJ | 3 | **0.83** | 4/4 | -55.6 | **0.51** | 4/4 | -57.1 | **-0.8** |
| 5′ P1, M-Box Riboswitch | 4 | 2QBZ | 1 | **0.96** | 4/4 | 126.1 | **0.61** | 4/4 | 126.1 | **-0.5** |
| 3′ J5/5a, Group I Intron | 4 | 2R8S | 1 | **0.47** | 4/4 | -21.4 | **0.40** | 4/4 | -21.4 | **0.0** |
| 5′ J5/5a, Group I Intron | 5 | 2R8S | 4 | **1.04** | 4/4 | -52.8 | **0.66** | 3/4 | -53.7 | **-0.9** |
| Hepatitis C Virus IRES IIa | 5 | 2PN4 | 1 | 5.31 | 0/0 | 22.1 | **0.71** | 0/0 | 22.1 | **-5.9** |
| J2/4, THP Riboswitch | 5 | 3D2V | 4 | **0.85** | 4/4 | -94.2 | **0.73** | 4/4 | -95.1 | **-1.0** |
| 23S rRNA (44-49) | 6 | 1S72 | 1 | **0.73** | 5/5 | -166.4 | **0.71** | 5/5 | -166.4 | **0.0** |
| 23S rRNA (531-536) | 6 | 1S72 | 5 | 2.45 | 3/4 | -260.3 | **0.76** | 3/4 | -261.3 | **-0.6** |
| J3/1, Glycine Riboswitch | 7 | 3OWI | 1 | **1.35** | 2/3 | 93.5 | **0.64** | 3/3 | 93.5 | 1.3 |
| J2/3, Group II Intron | 7 | 3G78 | 4 | **1.25** | 6/6 | -65.2 | **0.96** | 6/6 | -67.7 | **-0.2** |
| L1, SAM-II Riboswitch | 7 | 2QWY | 5 | **1.26** | 6/8 | 3.5 | **0.86** | 8/8 | 0.2 | **-1.3** |
| L2, Viral RNA Pseudoknot | 7 | 1L2X | 5 | 3.36 | 4/5 | -192.0 | **0.91** | 5/5 | -193.3 | **-4.1** |
| 23S rRNA (2534-2540) | 7 | 1S72 | 2 | 5.71 | 3/7 | -232.6 | **1.39** | 6/7 | -232.6 | **-7.3** |
| 23S rRNA (1976-1985) | 10 | 1S72 | 3 | 7.92 | 0/5 | -301.6 | 4.42 | 0/5 | -303.9 | **-7.5** |
| 23S rRNA (2003-2012) | 10 | 1S72 | 2 | **0.74** | 8/8 | -290.1 | **0.64** | 8/8 | -290.9 | 3.2 |
| AVERAGE | 6.3 | – | 2.8 | 2.28 | 3.8/4.7 | -99.1 | 0.99 | 4.2/4.7 | -100.1 | **-1.7** |
| **RMSD < 1.50 Å** | – | – | – | 10/15 | – | – | 14/15 | – | – | – |
| **Energy Gap <= 0.0** | – | – | – | – | – | – | – | – | – | 13/15 |

[a] Number of native base-pairs correctly recovered by the *de novo* model. Base pairs are annotated using an automated method based the scheme of Leontis and Westhof (RNA. 2001 Apr;7(4):499-512.) and recovery entails having the correct edge-to-edge interaction (Watson-Crick, Hoogsteen, or Sugar Edge) and local strand orientation (*cis* or *trans*). Counts of correctly recovered base pairs are lowered owing to ambiguities in assigning bifurcated base pairs and pairs connected by single hydrogen bonds.

[b] The Rosetta energy of the lowest energy native loop structure optimized through three different methods (see SI Supporting Methods for details). Bold text indicates that the lowest energy sampled by the *de novo* run is lower than the energy of the optimized native (i.e. the energy gap is negative).

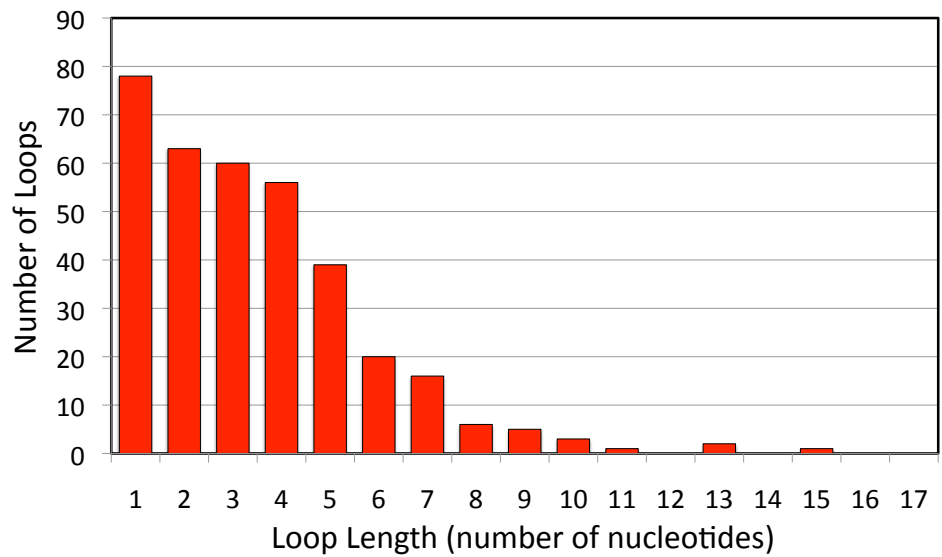

**Figure S1.** RNA loop length distribution in the 23S rRNA. A total of 350 single-stranded RNA loop segments separated by Watson-Crick and/or G-U wobble base pair(s) are found in the structure of the *Haloarcula marismortui* large ribosomal subunit (PDB: 1JJ2). The mean loop length is 3.5 nucleotides and loops longer than 10 nucleotides in length are rare (occur in <1%; 4 out of 350).

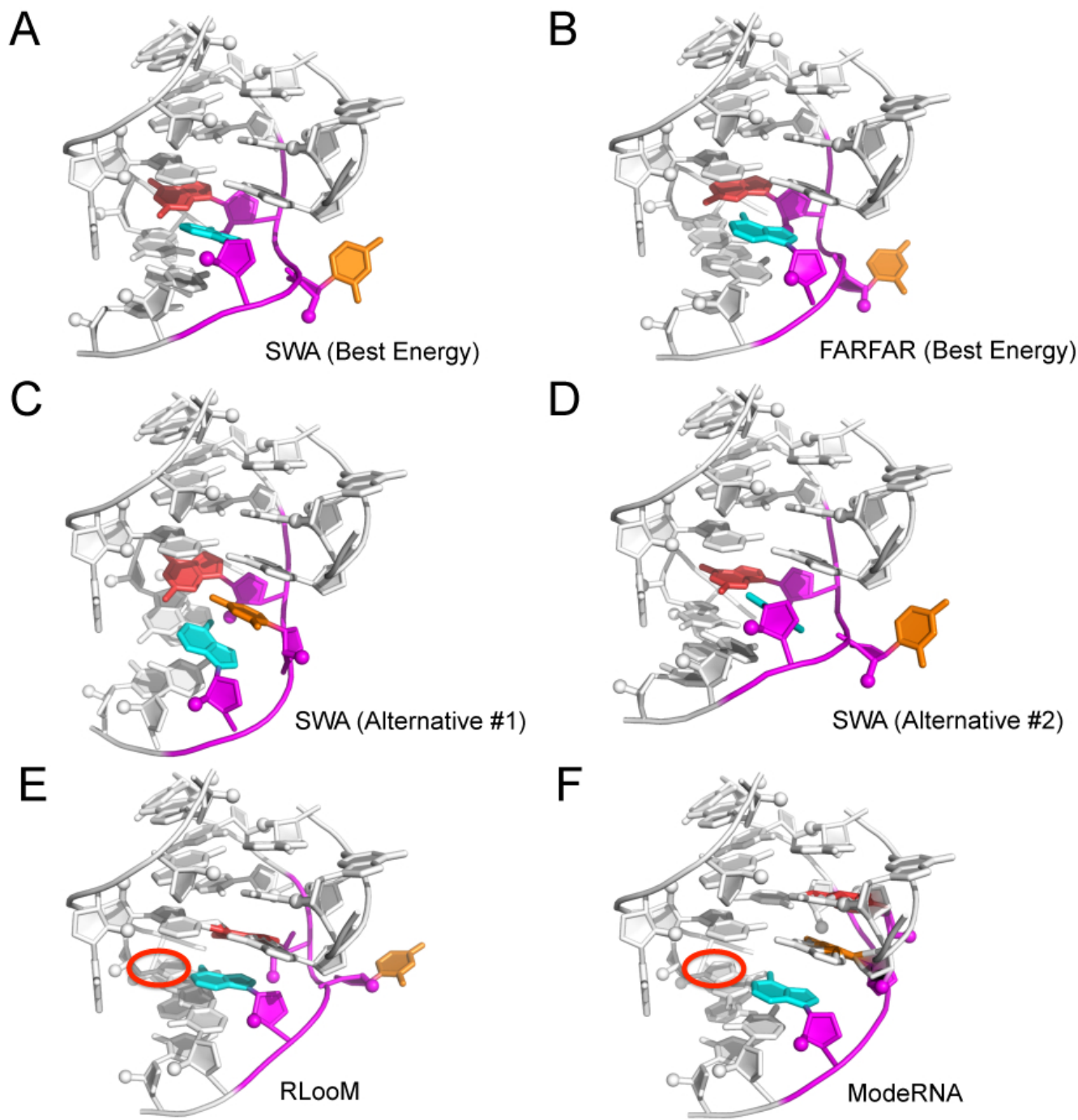

**Figure S2.** Comparsion of C7.2 Tetraloop Receptor model. (A) The lowest energy Stepwise Assembly model. Molprobity identified only 1 atom-atom clash in the model (excluding clashes inherited from the crystallographic model). (B) The lowest energy FARFAR model. Molprobity identified 3 atom-atom clashes in this model. (C and D) Alternative SWA models with poorer energies. In alternative model #1, U5 stacks between the $2^{nd}$ A of the tetraloop and A6. In alternative model #2, the base of A6 adopts a syn conformation and forms a *trans* sugar-Hoogsteen base pair with G4. (E) RLooM model. (F) ModeRNA model. Both the RLooM and the ModeRNA model are unlikely to be correct due to significant steric clashes (MolProbity identified 63 and 133 atom-atom clashes respectively) as well as an energetically unfavorable empty cavity in the core of the structure (red circle).

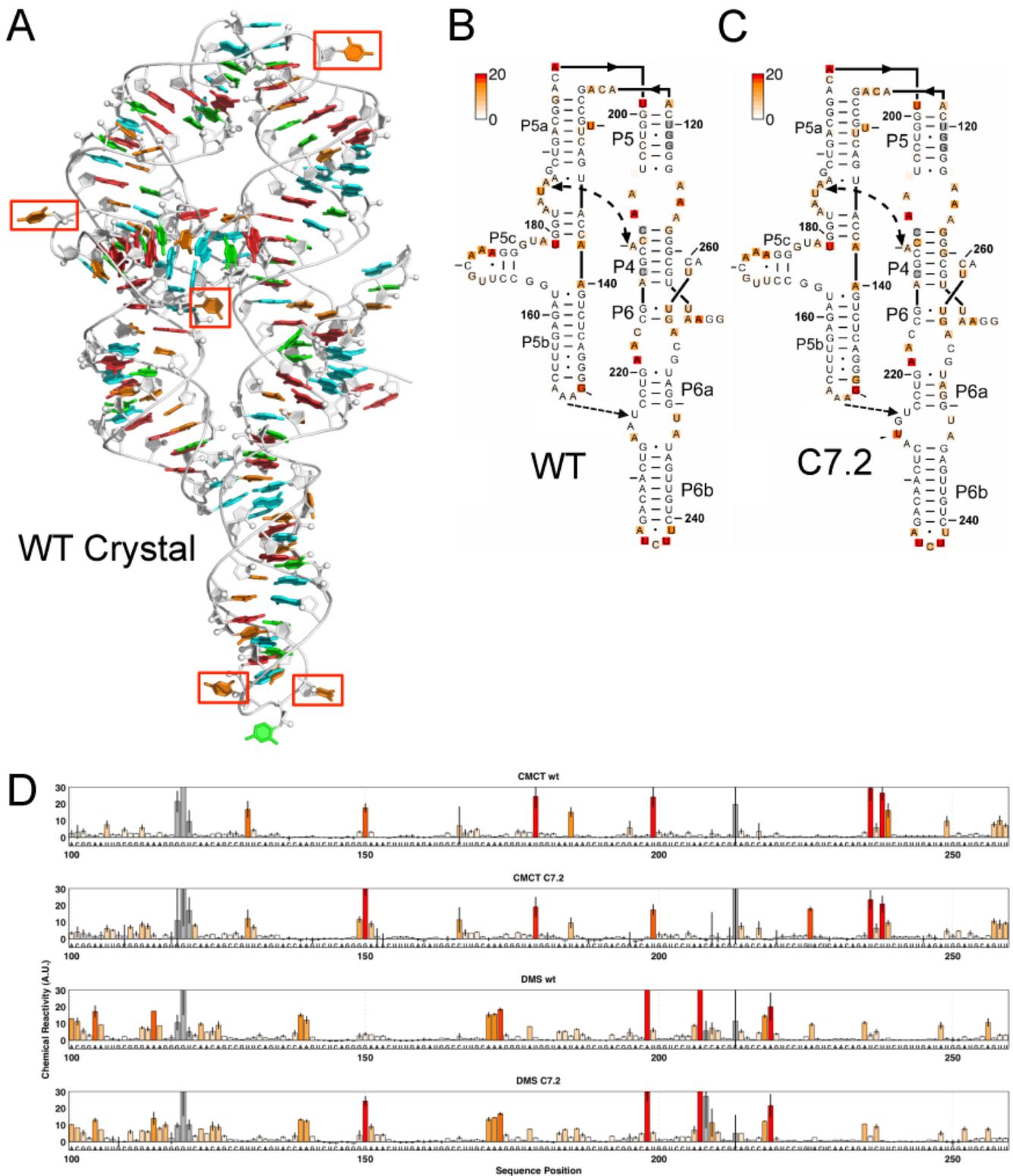

**Figure S3.** Chemical reactivity of the C7.2 tetraloop receptor, grafted into the P4-P6 RNA. (A) Crystallographic model of the P4-P6 domain of the *Tetrahymena* ribozyme (PDB: 2R8S), with all bulged uridines highlighted in red boxes. (B) Chemical reactivities of A and C (based on dimethyl sulfate alkylation) and G and U (based on CMCT carbodiimide modification) for the wild type P4-P6 RNA; measurements were acquired in 10 mM $MgCl_2$, 50 mM HEPES, pH 8.0, at 24 °C. Data were normalized so that reactivity in Watson-Crick base paired nucleotides average to unity. Nucleotides with high variability between replicates (greater than 10, due to high background stops) are colored in gray. (C) Chemical reactivities of the C7.2 mutant of the P4-P6 RNA. (D) DMS and CMCT reactivities plotted as separate bar graphs, with same coloring as (B) & (C). Error bars give standard deviations over four replicate measurements.

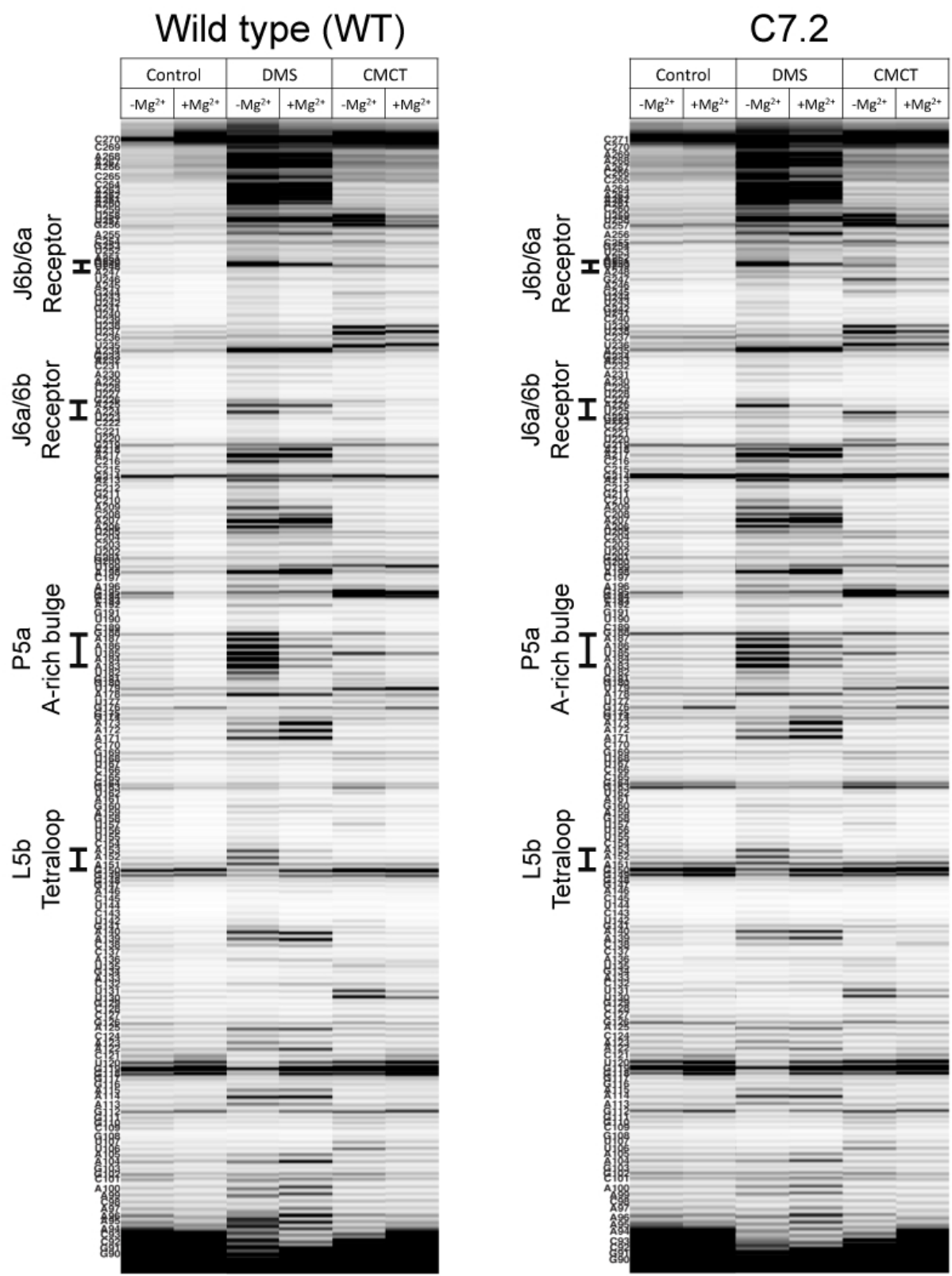

**Figure S4.** Capillary electropherograms of chemical mapping experiments on the wild type and C7.2-substituted P4-P6 RNAs. Modifications were read out by high throughput reverse transcription with 5′-fluorescently labeled radiolabeled primers and capillary electrophoresis. Raw fluorescence traces (arbitrary units) are shown after automated alignment and normalization to mean intensity. Shorter products (higher electrophoretic mobility) appear at the top. The 'Control' lanes contained no chemical modifier and provided background estimates for the measurements.

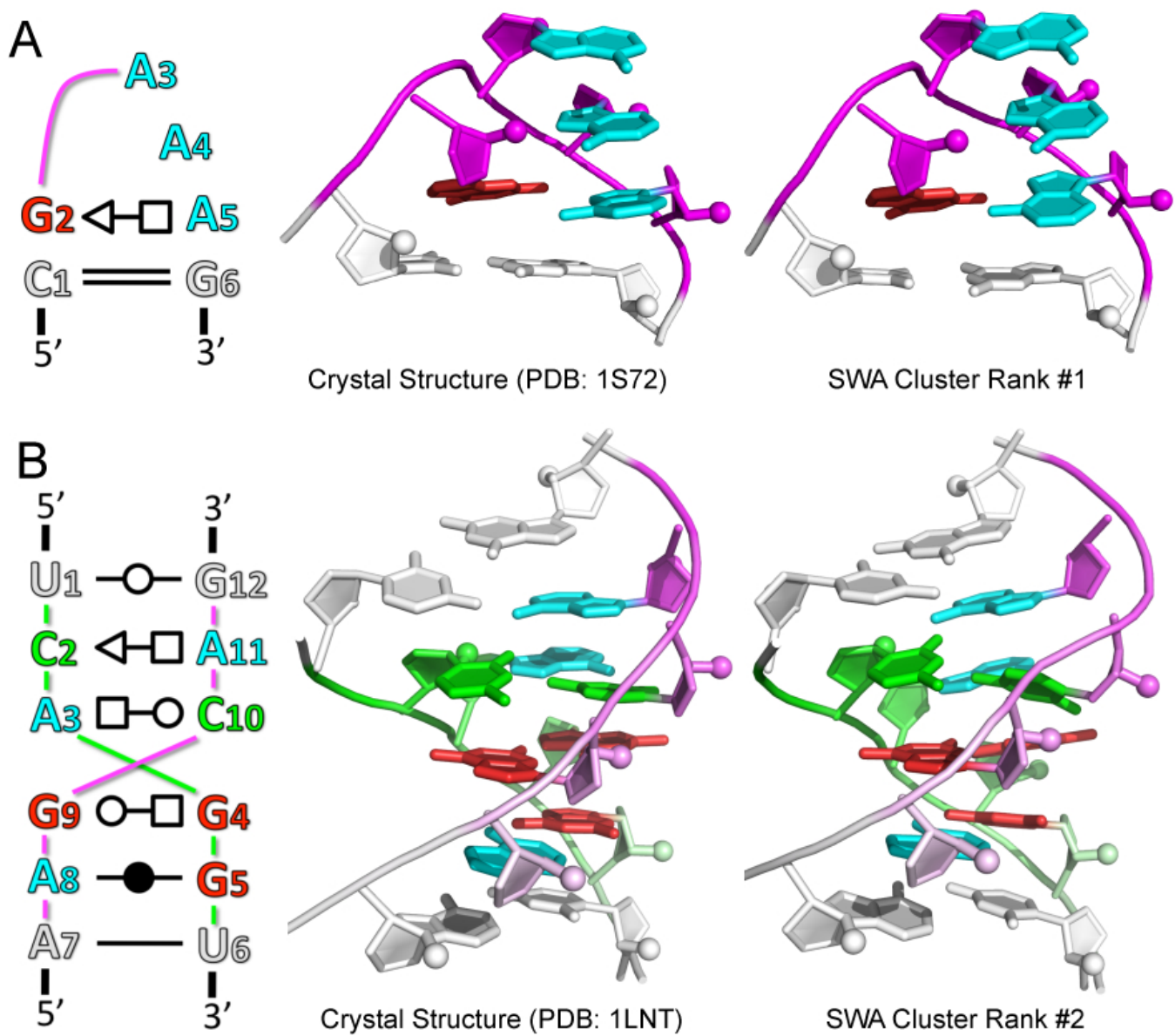

**Figure S5.** Extension of the SWA method to treat hairpins and multiple-stranded loops. SWA was able recapitulate the native conformation of (A) a 4-nt GAAA tetraloop hairpin (0.75 Å RMSD) and (B) a 8-nt double-stranded internal loop from the signal recognition particle RNA (0.98 Å RMSD). For the 4-nt GAAA tetraloop hairpin (PDB: 1S72), one correct building pathway is (C1-G6)→G2→A3→A4→A5 (Fig. S5A). For the 8-nt double-stranded internal loop from the signal recognition particle RNA (PDB: 1LNT), one correct building pathway is: (U1-G12)→C2→A11→C10→A3→G4→G9→G5→A8→(U6-A7) (Fig. S5B). Description of the extension of the SWA method to build hairpins and multiple-stranded loops *de novo* is provided in the SI Supporting Methods.